\documentclass[10pt,aps,prd,showpacs,twocolumn,longbibliography,nofootinbib,eqsecnum]{revtex4-2}
\usepackage[english]{babel}
\usepackage{amsfonts}
\usepackage{amssymb}
\usepackage{amsmath}
\usepackage{amsthm}
\usepackage{mathrsfs}
\usepackage{graphicx}
\usepackage{bbm}
\usepackage[unicode=true,bookmarksopen=true]{hyperref}
\usepackage{upgreek}
\usepackage{euscript}
\usepackage{color,soul}
\usepackage[normalem]{ulem}

\usepackage{etoolbox}
\apptocmd{\sloppy}{\hbadness 10000\relax}{}{}

\usepackage{mathtools}

\let\originalleft\left
\let\originalright\right
\renewcommand{\left}{\mathopen{}\mathclose\bgroup\originalleft}
\renewcommand{\right}{\aftergroup\egroup\originalright}

\makeatletter
\newenvironment{equations}[1][]{\subequations\ifx\relax#1\relax\else\label{#1}
\fi\align\ignorespaces}{\endalign\ignorespacesafterend\endsubequations}
\def\@spliteq#1{\begin{equation}\begin{split}#1\end{split}\end{equation}}
\def\splitequation{\collect@body\@spliteq}

\makeatother

\date{\today}

\newcommand{\eqend}[1]{\,#1}

\newcommand{\del}{\partial}

\newcommand{\beq}{\begin{equation}}
\newcommand{\eeq}{\end{equation}}
\newcommand{\beqa}{\begin{eqnarray}}
\newcommand{\eeqa}{\end{eqnarray}}
\newcommand{\cbar}{\bar{c}}
\newcommand{\rR}{\textrm{R}}
\newcommand{\rB}{\textrm{B}}

\begin{document}

\title{Infrared problem in the Faddeev-Popov ghost propagator in perturbative quantum 
gravity in de~Sitter spacetime}

\author{Jos Gibbons}
\email{josgibbons1988@gmail.com}
\affiliation{Department of Mathematics, University of York, Heslington, York YO10 5DD, United Kingdom}

\author{Atsushi Higuchi}
\email{atsushi.higuchi@york.ac.uk}
\affiliation{Department of Mathematics, University of York, Heslington, York YO10 5DD, United Kingdom}

\author{William~C.~C.~Lima}
\email{william.correadelima@york.ac.uk}
\affiliation{Department of Mathematics, University of York, Heslington, York YO10 5DD, United Kingdom}


\begin{abstract}
The propagators for the Faddeev-Popov (FP) ghosts in Yang-Mills theory and perturbative gravity in the covariant gauge are infrared (IR) divergent in de~Sitter spacetime.  An IR cutoff in the momentum space to regularize these divergences breaks the de~Sitter invariance. These IR divergences are due to the spatially constant modes in the Yang-Mills case and the modes proportional to the Killing vectors in the case of perturbative gravity. It has been proposed that these IR divergences can be removed, with the de~Sitter invariance preserved, by first regularizing them with an additional mass term for the FP ghosts and then taking the massless limit. In the Yang-Mills case, this procedure has been shown to correspond to requiring that the physical states, and the vacuum state in particular, be annihilated by some conserved charges in the Landau gauge. In this paper we show that there are similar conserved charges in perturbative gravity in the covariant Landau gauge in de~Sitter spacetime and that the IR-regularization procedure described above also correspond to requiring that the vacuum state be annihilated by these charges with a natural definition of the interacting vacuum state.
\end{abstract}

\maketitle

\section{Introduction}                                                                                                                              %
\label{sec:introduction}                                                                                                                            %

Inflationary cosmological models~\cite{Kazanas:1980tx,Sato:1980yn,Guth:1980zm,Linde:1981mu,Albrecht:1982wi} have been the main motivation for theoretical investigation into quantum field theory (QFT) in de~Sitter spacetime. The observation consistent with the assumption that the rate of expansion of our universe is accelerating~\cite{Riess:1998cb,Perlmutter:1998np} provides another motivation for this investigation. QFT in de~Sitter spacetime has been investigated also in the context of dS/CFT correspondence~\cite{Strominger:2001pn}.  Perturbative quantum gravity is not renormalizable, but it is still a theory with predictive power as an effective theory at each order of perturbation theory~\cite{Donoghue:1994dn}.

Perturbative quantum gravity in de~Sitter spacetime has many challenging features. Among them is the fact that the graviton propagator is infrared (IR) divergent in the physical gauge, with all gauge degrees of freedom fixed, natural to the spatially flat (or Poincar\'e) patch of this spacetime~\cite{Ford:1977dj}.  The source of the IR divergences is the similarity of graviton modes in this coordinate patch to those of massless minimally coupled scalar field~\cite{Ford:1977in,Allen:1985ux,Allen:1987tz}. However, it was found that these divergences do not manifest themselves in the physical quantities studied by the authors of Ref.~\cite{Ford:1977dj}. This finding is consistent with the fact that the IR-divergent part of the propagator can be written in pure-gauge form~\cite{Higuchi:2000ye,Allen:1986dd,Higuchi:2011vw}, i.e.\ that the IR-divergent part of the gravitational
perturbation can be expressed as $h_{\mu\nu} = \nabla_\mu A_\nu + \nabla_\nu A_\mu$.  (See Refs.~\cite{Frob:2017coq,Higuchi:2017sgj} for an analogous result for single field inflation.)  Some authors have claimed to show that these IR divergences would lead to breakdown of de~Sitter invariance (see, e.g.~Refs.~\cite{Tsamis:2011ep,Kitamoto:2012ep}), but this has not been established in a gauge-invariant manner.

The pure-gauge nature of the IR divergences in the sense explained above suggests that the graviton propagator may be IR finite in gauges natural to other coordinate patches. Indeed it is IR finite in the physical gauge natural to global coordinates of de~Sitter spacetime~\cite{Higuchi:2002sc}. Moreover, the covariant propagator in global coordinates is also IR finite~\cite{Faizal:2011iv}.  

Now, one also needs the Faddeev-Popov (FP) ghosts~\cite{Feynman:1963ax,Faddeev:1967fc,Faddeev:1973zb}, which are fermionic vector fields, in the covariant quantization of the gravitational field.  Although the graviton propagator is IR finite in global coordinates, the FP-ghost propagator is IR divergent. These IR divergences for the FP-ghost propagator are due to the modes proportional to the Killing vectors. However, the antighost field, $\bar{c}^\mu(x)$, appear in the Lagrangian density only in the form $\nabla_\mu \bar{c}_\nu + \nabla_\nu \bar{c}_\mu$, and for this reason the IR-divergent Killing-vector modes do not contribute to the interaction.  It has been proposed that the IR-divergences for the FP ghosts should be first regularized by the introduction of a small mass and that the massless limit should be taken at the end~\cite{Faizal:2008ns}. The resulting amplitude will be IR finite, i.e.\ it does not diverge in the massless limit, because the interaction terms are such that the IR divergences are eliminated because of the form of the interaction terms mentioned above. However, this procedure would appear rather \textit{ad hoc} and it needs further justification, in particular, with regards to its compatibility with the BRST invariance~\cite{Becchi:1975nq,Tyutin:1975qk}.

The FP-ghost propagator is IR divergent also in Yang-Mills theory in de~Sitter spacetime because the FP ghosts are massless minimally coupled scalar field in this theory. (In fact its propagator is IR divergent in any spacetime with compact Cauchy surfaces.) These IR divergences can also be regularized with a small mass term and by taking the massless limit at the end. The IR divergences in this case are due to the constant modes, and since only the derivatives of the antighost field appear in the interaction terms, the resulting amplitude is IR finite~\cite{Faizal:2008ns}.

For the Yang-Mills case, the procedure to eliminate the IR divergences from the FP-ghost sector mentioned above was shown to corresponds to requiring the vacuum state to be annihilated by certain conserved charges in the covariant Landau gauge~\cite{Gibbons:2014zya}. It was also proposed that all physical states be annihilated by these charges. (We note that a similar method has been used to eliminate the IR divergences in massless minimally coupled scalar field in de~Sitter spacetime~\cite{Kirsten:1993ug}.) These charges transform among themselves under BRST transformation, and hence 
this requirement on the vacuum state is compatible with, i.e.\ invariant under, the BRST transformation.

In this paper, we show that this equivalence holds also for perturbative quantum gravity in global de~Sitter spacetime in the covariant Landau gauge with a natural definition of the interacting vacuum state. That is, there are similar conserved charges in perturbative quantum gravity in this spacetime and the regularization and elimination of the IR divergences through a small mass term corresponds to  requiring that the vacuum state be annihilate by these charges in this gauge. These charges again transform among themselves under BRST transformation. Hence the requirement on the vacuum state is compatible with BRST invariance of the theory. Some of the results we present in the next sections were anticipated in Ref.~\cite{Gibbons_phd}.

The remainder of the paper is organized as follows. In Sec.~\ref{sec:symmetries} we present a brief description of de~Sitter spacetime, with emphasis on its Killing vectors.  In Sec.~\ref{sec:FPghostdivergence} we describe the IR divergences in the propagator of the FP ghosts for perturbative gravity in the covariant Landau gauge using the Euclidean formulation. In Sec.~\ref{sec:conserved-charges} we find the conserved charges which play the central role in this paper. In Sec.~\ref{sec:identify-Killing} we identify the conserved charges found in  Sec.~\ref{sec:conserved-charges} essentially as the canonical momenta conjugate to cyclic variables. Then, in Sec.~\ref{sec:Killing-vector-modes} we show that the regularization of the FP-ghost propagator with a mass term implies that the vacuum state is annihilated by the conserved charges found in Sec.~\ref{sec:conserved-charges} at tree level, i.e.\ in the free theory obtained by turning off the interaction. In Sec.~\ref{sec:Hamiltonian} we discuss our definition of the interacting vacuum state in Hamiltonian perturbation theory. This definition is combined with the result in the previous section to show that the interacting vacuum state is also annihilated by these charges. Finally in Sec.~\ref{sec:summary} we summarize and discuss our results. The Appendices contain some details omitted in the main text. Throughout this paper we employ units such that $G = \hbar = c = 1$ and adopt the signature $(-++\dots +)$ for the metric.

\section{Killing vectors in de~Sitter spacetime}                                                                                                    %
\label{sec:symmetries}                                                                                                                              %

In this section we discuss the Killing vectors in $n$-dimensional de~Sitter spacetime, which cause the IR divergences in the FP-ghost propagator. Consider $(n+1)$-dimensional Minkowski spacetime with Cartesian coordinates $X^\mu$, $\mu=0,1,\ldots,n$, and the metric
\beq
ds_{\textrm{M}}^2 = - (dX^0)^2 + \sum_{i=1}^n (dX^i)^2\eqend{.}  \label{original-metric}
\eeq
Then, the hypersurface defined by
\beq
-(X^0)^2 + \sum_{i=1}^n (X^i)^2 = 1/H^2\eqend{,}
\eeq
where $H$ is the Hubble constant, is the $n$-dimensional de~Sitter spacetime.  Let 
\begin{subequations}
\beqa
X^0 & = & H^{-1}\sinh Ht\eqend{,} \\
X^i & = & H^{-1}\cosh Ht\,\hat{x}^j,\ \ 1 \leq j \leq n\eqend{,}
\eeqa
\end{subequations}
where $t \in (-\infty,\infty)$ and $\sum_{i=1}^n (\hat{x}^j)^2 = 1$.  Thus, the coordinates $\hat{x}^j$ parametrize the unit $(n-1)$-sphere, $S^{n-1}$.  By substituting these formulas into Eq.~(\ref{original-metric}) we find the metric of de~Sitter spacetime as
\beq
ds^2 = - dt^2 + H^{-2}\cosh^2 Ht\, d\Omega_{n-1}^2\eqend{,}   \label{dS-metric}
\eeq
where $d\Omega_{n-1}^2$ is the metric on  $S^{n-1}$. From now on we let $H=1$ for simplicity.

The $n$-dimensional de~Sitter spacetime has the Killing symmetries of $(n+1)$-dimensional Minkowski spacetime with the origin fixed, i.e.\ $\mathfrak{so}(n,1)$.   There are $n(n-1)/2$ Killing vector fields generating the space rotations on $S^{n-1}$. In addition there are $n$ Killing vector fields generating the boosts in $n$ different directions.  These Killing symmetries are closely related to the IR divergences of the FP-ghost propagator as we find in the next section.

It is useful to remind ourselves of the spherical harmonics on $S^{n-1}$. The scalar spherical harmonics $Y_{(\ell\sigma)}(\boldsymbol{\theta})$, $\ell=0,1,2,\ldots$, on $S^{n-1}$, where $\boldsymbol{\theta}$ denotes the angular coordinates covering the sphere, satisfy~\cite{Rubin:1984tc}
\beq
\mathcal{D}_i \mathcal{D}^i Y_{(\ell\sigma)}(\boldsymbol{\theta}) = - \ell(\ell+n-2)Y_{(\ell\sigma)}(\boldsymbol{\theta})\eqend{,}\label{eq:scalar_spherical_harmonics}
\eeq
where the label $\sigma$ distinguishes between the scalar spherical harmonics with the same angular momentum $\ell$. Here, the covariant derivative $\mathcal{D}_i$ is compatible with the metric on $S^{n-1}$ and the indices are lowered and raised by the metric on $S^{n-1}$.
We require
\beq
\int_{S^{n-1}}d\Omega\, Y_{(\ell\sigma)}^*(\boldsymbol{\theta})Y_{(\ell'\sigma')}(\boldsymbol{\theta}) = \delta_{\ell\ell'}\delta_{\sigma\sigma'}\eqend{,}
\eeq
where $d\Omega$ is the surface element of $S^{n-1}$. The divergence-free vector spherical harmonics $Y_{(\ell\sigma)}^{i}(\boldsymbol{\theta})$ satisfy $\mathcal{D}_iY_{(\ell\sigma)}^i(\boldsymbol{\theta}) = 0$ and~\cite{Rubin:1984tc}
\beq
\mathcal{D}_k \mathcal{D}^k Y_{(\ell\sigma)}^i(\boldsymbol{\theta}) = \left[ -\ell(\ell+n-2)+1\right]Y_{(\ell\sigma)}^i(\boldsymbol{\theta})\eqend{.}
\eeq
We require
\beq
\int_{S^{n-1}}d\Omega\, Y_{(\ell\sigma)i}^*(\boldsymbol{\theta})Y_{(\ell'\sigma')}^i(\boldsymbol{\theta}) = \delta_{\ell\ell'}\delta_{\sigma\sigma'}\eqend{,}
\eeq
where the spatial index is lowered with the $S^{n - 1}$ metric.

The Killing vectors $\xi_{(\sigma,\textrm{R})}^\mu$ on de~Sitter spacetime that generate the rotations are given by 
\begin{subequations} \label{eq:rotation-Killing}
\beqa
\xi_{(\sigma,\textrm{R})}^{0} & = & 0\eqend{,}\\
\xi_{(\sigma,\textrm{R})}^{i} & =  & Y_{(1\sigma)}^{i}\eqend{,}
\eeqa
\end{subequations}
 i.e.\
$Y_{(\ell\sigma)}^{i}$ with $\ell=1$.  The Killing vectors $\xi_{(\sigma,\textrm{B})}^\mu$ on de~Sitter spacetime that generate the boosts are given by 
\begin{subequations} \label{eq:boost-Killing}
\beqa
\xi_{(\sigma,\textrm{B})}^0 & = & Y_{(1\sigma)}\eqend{,}\\
\xi_{(\sigma,\textrm{B})}^i & = &  \tanh t\,\mathcal{D}^i Y_{(1\sigma)}\eqend{,}
\eeqa
\end{subequations}
where the index $i$ is raised by the metric on $S^{n-1}$.

The metric \eqref{dS-metric} on de~Sitter spacetime (with $H=1$) becomes that of the unit $n$-sphere ($S^n$),
\beq
d\Omega_n^2 = d\tau^2 + \sin^2\tau d\Omega_{n-1}^2\eqend{,}
\eeq
by the following complex coordinate transformation:
\beq
\tau = \frac{\pi}{2} + it\eqend{.}  \label{eq:Euclidean-time}
\eeq
Upon this coordinate transformation, both types of Killing vectors $\xi_{(\sigma,\textrm{R})}^\mu$ and $\xi_{(\sigma,\textrm{B})}^\mu$ become, up to
constant normalization factors, the rotation Killing vectors $V^\mu_{(1\rho)}$, where $V^\mu_{(L\rho)}$, $L=1,2,\ldots$, are the divergence-free vector spherical harmonics on $S^{n}$, satisfying the eigenvalue equation,
\beq
\nabla_\nu \nabla^\nu V^\mu_{(L\rho)}(\tau,\boldsymbol{\theta}) = \left[ - L(L+n-1)+1\right] V^\mu_{(L\rho)}(\tau,\boldsymbol{\theta})\eqend{,} \label{V-mu-eq1}
\eeq
and the normalization condition
\beq
\int_{S^{n}}d\Omega\, V_{(L\sigma)\mu}^*(\tau,\boldsymbol{\theta})V_{(L'\rho')}^\mu(\tau,\boldsymbol{\theta}) = \delta_{LL'}\delta_{\rho\rho'}\eqend{,}
\label{V-mu-eq2}
\eeq
where $d\Omega$ is the volume element on $S^n$.

\section{IR divergences in  the FP-ghost propagator}                                                                                                %
\label{sec:FPghostdivergence}                                                                                                                       %

In this section we discuss the structure of the IR divergences of the FP-ghost propagator.  The Lagrangian density for perturbative gravity in the covariant Landau gauge reads
\beq
\mathcal{L} = \mathcal{L}_{\textrm{GR}} + \sqrt{-g}\mathcal{L}_{\textrm{FP}} +
\sqrt{-g}\mathcal{L}_{\textrm{gf}}\eqend{,}
\eeq
where $\mathcal{L}_{\textrm{GR}}$ is the diffeomorphism invariant Lagrangian density describing the gravitational field  and where $g$ is the determinant of the \textit{background} metric tensor $g_{\mu\nu}$. The gauge-fixing and FP-ghost Lagrangian densities\footnote{In this paper the quantity obtained by dividing a Lagrangian density by $\sqrt{-g}$ is also called a Lagrangian density.}  are given by
\begin{subequations} \label{eq:both-FP-and-gf}
\beqa
\mathcal{L}_{\textrm{gf}} & = & - \nabla^\mu B^\nu (h_{\mu\nu} - k g_{\mu\nu} {h^\alpha}_\alpha)\eqend{,}
\label{eq:gf-Lag} \\
\mathcal{L}_{\textrm{FP}} 
& = & - i \nabla^\mu \bar{c}^\nu (\nabla_\mu c_\nu + \nabla_\nu c_\mu
- 2k g_{\mu\nu} \nabla_\alpha c^\alpha \nonumber \\
&& \ \ \ \ \  \ \ \ \ \ \ \ \ \ + \pounds_c h_{\mu\nu} - k 
g_{\mu\nu}g^{\alpha\beta}\pounds_c h_{\alpha\beta}), \label{FP-Lag}
\eeqa
\end{subequations}
where $\pounds_X$ denotes the Lie derivative in the direction of the vector $X^\mu$.  That is,
\beq
\pounds_c h_{\mu\nu} = c^\alpha\nabla_\alpha h_{\mu\nu} + (\nabla_\mu c^\alpha)h_{\alpha\nu}
+ (\nabla_\nu c^\alpha)h_{\mu\alpha}\eqend{.} \label{Lie_of_h_mu_nu}
\eeq
The field $h_{\mu\nu}$ is the gravitational perturbation: the full metric is given by $\tilde{g}_{\mu\nu} = g_{\mu\nu} + h_{\mu\nu}$, where $g_{\mu\nu}$ is the background de~Sitter metric.  The indices in Eq.~(\ref{eq:both-FP-and-gf}) are raised and lowered by $g_{\mu\nu}$.  The ghost and antighost fields $c^\mu$ and $\bar{c}^\mu$, respectively, are anticommuting Hermitian fields~\cite{Kugo:1977zq,Kugo:1979gm}.  

The gauge-fixing term would fail to provide a time-derivative of $h_{00}$ if $k=1$.  This value is excluded for this reason, and it is often convenient to write
\beq\label{eq:beta}
k = 1 + \frac{1}{\beta} \eqend{.} 
\eeq
The parameter $\beta$ will be taken to be real, but outside the set $-s(s + n - 1)/(n - 1)$, with $s = 0,1,2,\dots$. The gauge-fixing Lagrangian density $\mathcal{L}_{\textrm{gf}}$ is the $\alpha\to 0$ limit of 
\beq
\mathcal{L}^{(\alpha)}_{\textrm{gf}} = \frac{\alpha}{2} 
B^\mu B_\mu - \nabla^\mu B^\nu H_{\mu\nu}\eqend{,}
\eeq
where we have defined
\beq
H_{\mu\nu} \equiv h_{\mu\nu} - k g_{\mu\nu}h \eqend{,}  \label{eq:def-of-Hmunu}
\eeq
with $h \equiv {h^\alpha}_\alpha$.
By defining 
\begin{equation}
\hat{B}^\mu \equiv B^\mu + \frac{1}{\alpha}\nabla_\nu H^{\mu\nu}
\end{equation}
and neglecting total-divergence terms, we have
\beq
\mathcal{L}^{(\alpha)}_{\textrm{gf}} = \frac{\alpha}{2} \hat{B}^\mu \hat{B}_\mu - \frac{1}{2\alpha}\nabla^\nu H_{\mu\nu}
\nabla_\lambda H^{\mu\lambda}\eqend{.} \label{eq:new-gf-term}
\eeq 
The field $\hat{B}^\mu$ can be neglected because it is decoupled from other fields. The remaining term is the gauge-fixing term more commonly
used. Note that the Euler-Lagrange equation from varying $B^\mu$ in the Lagrangian density $\mathcal{L}_{\textrm{gf}}$ reads
\beq
\nabla_\nu H^{\mu\nu} = 0\eqend{.}
\eeq
Thus, in this gauge the gauge condition is a result of a field equation. 

The FP-ghost Lagrangian density~\eqref{FP-Lag} was constructed so that $\mathcal{L}_{\textrm{gf}} + \mathcal{L}_{\textrm{FP}}$ is BRST invariant, following the well-known general procedure (see, e.g., Ref.~\cite{Kugo:1981hm}).  The BRST transformation is given as follows:
\begin{subequations} \label{eq:BRST-all}
\beqa
\delta_{\textrm{B}} h_{\mu\nu} & = & 
\nabla_\mu c_\nu + \nabla_\nu c_\mu + \pounds_c h_{\mu\nu}\eqend{,}\label{BRSTtr1}\\
\delta_{\textrm{B}}c^\mu & =  & c^\alpha\nabla_\alpha c^\mu\eqend{,}\\
\delta_{\textrm{B}}\bar{c}^\mu & = & i B^\mu\eqend{,}\\
\delta_{\textrm{B}}B^\mu & = & 0\eqend{,}\label{BRSTtr4}
\eeqa
\end{subequations}
where $\pounds_c h_{\mu\nu}$ is given by Eq.~\eqref{Lie_of_h_mu_nu}. The transform $\delta_{\textrm{B}}h_{\mu\nu}$ can be understood as the Lie derivative with respect to $c^\mu$ of the full metric $\tilde{g}_{\mu\nu} = g_{\mu\nu} + h_{\mu\nu}$.  Thus, the action for the gravitational field obtained by integrating $\mathcal{L}_{\textrm{GR}}$ over the spacetime is invariant under this transformation.  It can readily be verified that the BRST transformation given by Eq.~\eqref{eq:BRST-all} is nilpotent, i.e.\ $\delta_{\textrm{B}}^2 = 0$~\cite{Kugo:1979gm}. Indeed, we find that
\beqa
\delta_{\textrm{B}}^2c^\mu & = & (\delta_{\textrm{B}}c^\alpha)\nabla_\alpha c^\mu 
- c^\alpha \nabla_\alpha \delta_{\textrm{B}}c^\mu \nonumber \\
& = & - {R^\mu}_{\alpha\beta \gamma}c^\alpha c^\beta c^\gamma \nonumber \\
& = & 0\eqend{,}
\eeqa
where we have defined $\delta_{\textrm{B}}$ to act from the left so that
$\delta_{\textrm{B}} (\Omega_1\Omega_2) = (\delta_{\textrm{B}}\Omega_1)\Omega_2 - \Omega_1 
\delta_{\textrm{B}}\Omega_2$ if 
$\Omega_1$ is fermionic. 
The equality $\delta_{\textrm{B}}^2h_{\mu\nu} = 0$ follows from
\beq
\pounds_c \pounds_c \tilde{g}_{\mu\nu} = \pounds_{X} \tilde{g}_{\mu\nu}\eqend{,}
\eeq
where $X^\mu \equiv c^\alpha \nabla_\alpha c^\mu$.  It can readily be seen that
\beq
\mathcal{L}_{\textrm{FP}} + \mathcal{L}_{\textrm{gf}}
= i \delta_{\textrm{B}}[(\nabla^\mu \bar{c}^\nu)H_{\mu\nu}]\eqend{.}
\eeq
The BRST invariance of $\mathcal{L}_{\textrm{FP}} + \mathcal{L}_{\textrm{gf}}$ follows from the nilpotency of $\delta_{\textrm{B}}$.

Now, let us discuss the IR divergences in the FP-ghost propagator.  The free field equation, i.e.\ the equation obtained by dropping the interaction terms, for the ghost field is
\beq
\nabla^\nu (\nabla_\mu c_\nu + \nabla_\nu c_\mu - k g_{\mu\nu}\nabla_\alpha c^{\alpha})
= 0\eqend{.} \label{eq:c-free-field}
\eeq
From here to the end of this section, the fields $c^\mu$ and $\cbar^\mu$ are assumed to satisfy the free field equation. The free antighost field $\bar{c}^{\mu}$ satisfies the same equation. It is convenient to rewrite Eq.~\eqref{eq:c-free-field} by interchanging some derivatives and 
using $R_{\mu\nu} = (n-1)g_{\mu\nu}$ as
\beq
{L_\mu}^\nu c_\nu = 0\eqend{,}
\eeq
where the differential operator ${L_\mu}^\nu$ is given by 
\begin{splitequation}
{L_\mu}^\nu 
& = - \delta_\mu^\nu \nabla_\alpha\nabla^\alpha + \nabla^\nu \nabla_\mu + 2\beta^{-1} \nabla_\mu \nabla^\nu \\
&\phantom{=}\; - 2(n-1)\delta_\mu^\nu + m^2\delta_\mu^\nu \eqend{.}
\end{splitequation}
We have inserted a mass term $m^2 \delta_\mu^\nu$ as an IR regulator.

By writing the tree-level Feynman propagator as
\beq
G^{(\textrm{FP})}_{\mu\mu'}(x,x') \equiv -iT \langle 0| c_\mu(x) \bar{c}_{\mu'}(x')|0\rangle\eqend{,}
\eeq
one finds that the function $G^{(\textrm{FP})}_{\mu\mu'}(x,x')$ satisfies
\beq
{L_\mu}^\nu G^{(\textrm{FP})}_{\nu\nu'}(x,x') = g_{\mu\nu'} \delta^{(n)}(x,x')\eqend{,} \label{field-eq-for-c-prop}
\eeq
where the delta function $\delta^{(n)}$ is defined to have the property
\beq
\int d^n x\,\sqrt{-g(x)}\, f(x)\delta^{(n)}(x,x') = f(x')\eqend{,}
\eeq
for any compactly-supported smooth function $f(x)$.  The differential operator ${L_\mu}^\nu$ acts on $x$ in Eq.~(\ref{field-eq-for-c-prop}).

The IR divergences of the FP-ghost propagator in the Bunch-Davies (or Euclidean) vacuum state~\cite{Chernikov:1968zm,Schomblond:1976xc,Gibbons:1977mu,Bunch:1978yq} in the $n$-dimensional de~Sitter background is best understood in the Euclidean approach.  The Feynman propagator $G^{(\textrm{FP})}_{\mu\mu'}(x,x')$ on de~Sitter spacetime in the Euclidean vacuum state can be obtained by finding the (unique) solution to Eq.~(\ref{field-eq-for-c-prop}) on the $n$-dimensional sphere, $S^n$, and then analytically continuing it to de~Sitter spacetime by the relation~\eqref{eq:Euclidean-time}. Since Eq.~(\ref{field-eq-for-c-prop}) shows that $G^{(\textrm{FP})}_{\mu\mu'}(x,x')$ is the inverse of the differential operator ${L_\mu}^\nu$, it can be expressed in terms of the eigenfunctions of this differential operator on $S^{n}$. Any smooth vector field on $S^{n}$ can be expressed as a linear combination of the divergence-free vector eigenfunctions $V_{(L\rho)}^\mu$ discussed in the previous section and the gradient eigenfunctions $\nabla_\mu \phi_{(L\rho)}$, $L=1,2,3,\ldots$ of the Laplace-Beltrami operator $\nabla_\nu \nabla^\nu$. The vectors $V_{(L\rho)}^\mu$ satisfy Eqs.~\eqref{V-mu-eq1} and \eqref{V-mu-eq2} whereas the functions
$\phi_{(L\rho)}$ satisfy
\beq
\nabla^\nu \nabla_\nu \phi_{(L\rho)} =  - L(L+n-1)\phi_{(L\rho)}\eqend{.}
\eeq
It is convenient to normalize these functions as follows:
\beq
\int_{S^n} dS\, \phi_{(L\rho)}^*\phi_{(L'\rho)}  =  \frac{1}{L(L+n-1)}
\delta_{LL'}\delta_{\rho\rho'}\eqend{.} 
\eeq
One readily finds
\begin{subequations}
\beqa
{L_\mu}^\nu V_{(L\rho)\nu} & = & \left[(L-1)(L+n)+m^2\right] V_{(L\sigma)\mu}\eqend{,} \\
{L_\mu}^\nu \nabla_\nu \phi_{(L\rho)} & = & \left[-2\beta^{-1}L(L+n-1) \right. \nonumber \\
&& \ \ \ \ \ \ \ \ \ \left. - 2(n-1) + m^2\right]\nabla_\mu \phi_{(L\rho)}
\eqend{.}\nonumber \\
\eeqa 
\end{subequations}
Hence,
\begin{splitequation}
& G^{(\textrm{FP})}_{\mu\mu'}(x,x') \\
& \; = \sum_{L=1}^\infty \sum_\sigma \frac{V_{(L\sigma)\mu}(x)V_{(L\sigma)\mu'}^*(x')}{(L-1)(L+n)+m^2}\\
&\; \phantom{=}\; - \beta\sum_{L=1}^\infty \sum_{\sigma} \frac{\nabla_\mu\phi_{(L\sigma)}(x)\nabla_{\mu'}\phi_{(L\sigma)*}(x')}{2\left\{L(L+n-1)+(n-1)\beta\right\} - m^2}\\
& \; =  \frac{1}{m^2}\sum_{\sigma} V_{(1\sigma)\mu}(x)V_{(1\sigma)\mu'}^*(x') + G_{\mu\mu'}^{(\textrm{FP},\textrm{reg})}(x,x')\eqend{,}
\end{splitequation}
where the function $G_{\mu\mu'}^{(\textrm{FP},\textrm{reg})}(x,x')$ remains finite in the limit $m\to 0$. (Recall that $\beta > 0$.)

Now, as we saw in the previous section, the vectors $V_{(1\sigma)}^{\mu}$ are the Killing vectors on $S^{n}$. Hence, upon analytic continuation, 
one finds that the Feynman propagator for the FP ghosts takes the form
\beqa
G^{(\textrm{FP})}_{\mu\mu'}(x,x')
& = & \frac{1}{m^2}\sum_{A} c_A\xi_{A\mu} (x)\xi_{A\mu'}(x') \nonumber \\
&& + G_{\mu\mu'}^{(\textrm{FP},\textrm{reg})}(x,x')\eqend{,}
\eeqa
where $\xi_{A}^\mu(x)$ with $A=(\sigma,\textrm{R})$ or  $(\sigma,\textrm{B})$ are the Killing vectors in de~Sitter spacetime and $c_A$ are constants.
Although the Feynman propagator $G^{(\textrm{FP})}_{\mu\mu'}(x,x')$ is IR divergent, i.e.\ it diverges as $m\to 0$, if one uses the regularized propagator in perturbative calculations, the IR divergences cancel out.  This is because all interaction terms in $\mathcal{L}_{\textrm{FP}}$ involve the factor $\nabla_\mu \bar{c}_\nu + \nabla_\nu \bar{c}_\mu$. Indeed, from the Killing equation $\nabla_\mu \xi_{A\nu} + \nabla_\nu \xi_{A\mu} = 0$, one finds
\beqa
&& \nabla_{\mu'} G^{(\textrm{FP})}_{\mu\nu'}(x,x') + \nabla_{\nu'}G^{(\textrm{FP})}_{\mu\mu'}(x,x') \nonumber\\
&& = \nabla_{\mu'} G^{(\textrm{FP},\textrm{reg})}_{\mu\nu'}(x,x') 
+ \nabla_{\nu'}G^{(\textrm{FP},\textrm{reg})}_{\mu\mu'}(x,x')\eqend{,}
\eeqa
where the derivative $\nabla_{\mu'}$ acts on $x'$.  Thus, the use of the regularized FP-ghost propagator will lead to IR-finite amplitudes.

For this reason, it was proposed in Ref.~\cite{Faizal:2008ns} that one should use the regularized FP-ghost propagator and take the massless limit after the calculation, thus preserving the de~Sitter invariance, rather than breaking it by introducing a momentum cutoff. However, since a mass term breaks the BRST invariance, it was not clear whether such a procedure leads to a consistent theory.  The purpose of this paper is to establish that the use of the regularized FP-ghost propagator corresponds to requiring that the vacuum state be annihilated by certain conserved charges in a BRST-invariant manner and, hence, that such a procedure is consistent with the BRST invariance of the theory.  This equivalence is an analogue of that for Yang-Mills theory in de~Sitter spacetime demonstrated in Ref.~\cite{Gibbons:2014zya}.

\section{Conserved charges in perturbative gravity in the Landau gauge}                                                                             %
\label{sec:conserved-charges}                                                                                                                       %

In this section we find some conserved charges in perturbative gravity in the Landau gauge about a background spacetime satisfying Einstein's equations with compact Cauchy surfaces and with Killing symmetries.  Such spacetimes include global de~Sitter spacetime. We also show how these charges are related to one another by BRST transformation.
 
First, the field equation $\nabla_\nu H^{\mu\nu} = 0$, where $H_{\mu\nu}$ is defined by Eq.~\eqref{eq:def-of-Hmunu}, and the Killing equation 
$\nabla_\mu \xi_{A\nu} + \nabla_\nu \xi_{A\mu} =0$ imply that $\nabla_\mu (\xi_{A\nu} H^{\mu\nu}) = 0$.  Hence the following charges are conserved:
\beqa
Q_{A}^{(H)} & \equiv & \int_{\Sigma} d\Sigma\, n_\mu \xi_{A\nu} H^{\mu\nu}\eqend{,}\label{eq:bosonic-charge}
\eeqa
where $n^\mu$ is the future-pointing unit normal to the Cauchy surface $\Sigma$, i.e.\ $n^0 > 0$ and $n_\mu n^\mu = -1$, and where $d\Sigma$ is the hypersurface element on $\Sigma$.

Next, we note that the field equation coming from varying $\bar{c}^\mu$ is also the divergence of a symmetric tensor, i.e.
\beq
\nabla_\mu \mathcal{S}^{\mu\nu} = 0\eqend{,}
\eeq
where
\beqa
\mathcal{S}_{\mu\nu}
 & =&  \nabla_\mu c_\nu + \nabla_\nu c_\mu - 2k g_{\mu\nu}\nabla_\alpha c^\alpha\nonumber \\
&& + \pounds_c h_{\mu\nu} - kg_{\mu\nu} g^{\alpha\beta}\pounds_{c} h_{\alpha\beta}\eqend{.}
\label{mathcalC}
\eeqa
Then, $\nabla_\mu (\xi_{A\nu}\mathcal{S}^{\mu\nu}) = 0$.  Hence, the charges given by
\beq
Q_A^{(c)} \equiv \int_\Sigma d\Sigma\,n_\mu \xi_{A\nu}\mathcal{S}^{\mu\nu}
\eeq
are conserved.  It is clear that $Q_A^{(c)} = \delta_\textrm{B} Q_A^{(H)}$ since $\mathcal{S}_{\mu\nu} = \delta_\textrm{B} H_{\mu\nu}$.

It is convenient to remove the derivatives on $h_{\mu\nu}$ in the expression for $Q_A^{(c)}$ for later purposes. We observe
\begin{splitequation}
\xi_{A\nu} \mathcal{S}^{\mu\nu} 
& = \xi_{A\nu}\widetilde{\mathcal{S}}^{\mu\nu} - (\nabla_\alpha \xi_{A\nu})c^\alpha H^{\mu\nu}\\ 
& \phantom{=} + 2\nabla_\alpha(\xi_{A\nu}c^{[\alpha} H^{\mu]\nu})\eqend{,}
\end{splitequation}
where
\begin{splitequation}
\widetilde{\mathcal{S}}_{\mu\nu} 
& = (\nabla_\mu c_\nu + \nabla_\nu c_\mu - 2kg_{\mu\nu} \nabla_\alpha c^\alpha)(1+kh) \\
& \phantom{=} + (\nabla^\alpha c_\mu) H_{\alpha\nu} - (\nabla_\alpha c^\alpha) H_{\mu\nu} \\
& \phantom{=} + (\nabla_\mu c^\alpha)H_{\alpha\nu} + (\nabla_\nu c^\alpha) H_{\alpha\mu}\\ 
&\phantom{=} - 2k g_{\mu\nu} (\nabla^\alpha c^\beta)H_{\alpha\beta}\eqend{.}
\label{eq:tildeS}
\end{splitequation}
Then these charges can be expressed as follows:
\beq
Q_A^{(c)} = \int_\Sigma d\Sigma\,n_\mu \left[
\xi_{A\nu}\widetilde{\mathcal{S}}^{\mu\nu} - (\nabla_\alpha \xi_{A\nu})c^\alpha H^{\mu\nu}\right]\eqend{,}
\label{eq:charge-c}
\eeq
because the integral of a vector of the form $\nabla_\mu F^{\mu\nu}$, where $F^{\mu\nu}$ is an antisymmetric tensor, over any (compact) Cauchy surface vanishes by the generalized Stokes theorem.

There are also conserved charges arising from the field equation coming from varying $c^\mu$.  To show this, it is convenient to write the sum of the FP and gauge-fixing Lagrangian densities in the following form:
\begin{splitequation}
& \mathcal{L}_{\textrm{FP}} + \mathcal{L}_{\textrm{gf}} \\
&\; = -i (\nabla_\mu \bar{c}_\nu + \nabla_\nu \bar{c}_\mu - 2k g_{\mu\nu}\nabla_\alpha \bar{c}^\alpha + \pounds_{\bar{c}}h_{\mu\nu} \\
&\; \phantom{=}\; - k g_{\mu\nu} g^{\beta\gamma}\pounds_{\bar{c}}h_{\beta\gamma}) \nabla^\mu c^{\nu}  - \nabla^\mu \bar{B}^\nu H_{\mu\nu}\\
&\; \phantom{=}\; + i (1-2k)\left(\nabla_\mu \cbar_\nu \nabla_\alpha c^\alpha - \nabla_\alpha\cbar^\alpha \nabla_\mu c_\nu\right) H^{\mu\nu}\eqend{,}
\label{newlagrangian}
\end{splitequation}
where
\begin{equation}
\bar{B}^\nu \equiv B^\nu + i\left[ \cbar^\alpha \nabla_\alpha c^\nu - (\nabla_\alpha \cbar^\nu)c^\alpha\right]\eqend{.}
\label{B-def}
\end{equation}
Equation \eqref{newlagrangian} is derived in Appendix~\ref{appA}. Notice that, if $k=1/2$, then this equation will be equal to the negative of the original one with $B^\mu$ replaced by $-\bar{B}^\mu$ and with $c^\mu$ and $\bar{c}^\mu$ interchanged.   The field equation arising from  
varying the Lagrangian density \eqref{newlagrangian} with respect to $c^\mu$ is then
\beq
\nabla_\nu \mathcal{T}^{\mu\nu} = 0\eqend{,}
\eeq
where
\begin{eqnarray}
\mathcal{T}_{\mu\nu}
&  = &   \nabla_\mu \cbar_\nu  + \nabla_\nu \cbar_\mu - 2k g_{\mu\nu}\nabla_\alpha \cbar^\alpha \nonumber \\
&&  + \pounds_{\bar{c}}h_{\mu\nu}
- k g_{\mu\nu}g^{\beta\gamma}\pounds_{\bar{c}}h_{\beta\gamma} \nonumber \\
&& + (2k-1)(g_{\mu\nu} \nabla^\beta \bar{c}^\gamma h_{\beta\gamma} 
- h_{\mu\nu} \nabla_\alpha \bar{c}^\alpha)\eqend{.}
\label{eq:Tmunu}
\end{eqnarray}
The contribution from varying $c^\mu$ through $\bar{B}^\mu$ in Eq.~\eqref{B-def} has not been included here because it is proportional to 
$\nabla_\nu H^{\mu\nu}$, which vanishes by a field equation. Then, since $\nabla_\mu (\xi_{A\nu} \mathcal{T}^{\mu\nu}) = 0$ for any Killing vector $\xi_{A}^\mu$, we have the following conserved charges:
\beq
Q_A^{(\cbar)} \equiv \int_\Sigma d\Sigma\,n_\mu \xi_{A\nu}\mathcal{T}^{\mu\nu}\eqend{.}
\label{charge-cbar}
\eeq
The similarity of $\mathcal{S}^{\mu\nu}$ and $\mathcal{T}^{\mu\nu}$ allows us to use the same method to remove the derivative of $h_{\mu\nu}$ from these charges.  Thus, by defining
\beqa
\widetilde{\mathcal{T}}_{\mu\nu} 
& = & (\nabla_\mu \cbar_\nu + \nabla_\nu \cbar_\mu - 2kg_{\mu\nu} \nabla_\alpha \cbar^\alpha)(1+kh) \nonumber\\
&& + (\nabla^\alpha \cbar_\mu) H_{\alpha\nu}
- 2k (\nabla_\alpha \cbar^\alpha) H_{\mu\nu} \nonumber \\
&& + (\nabla_\mu \cbar^\alpha)H_{\alpha\nu} + (\nabla_\nu \cbar^\alpha)
H_{\alpha\mu} - g_{\mu\nu} (\nabla^\alpha \cbar^\beta)H_{\alpha\beta}\eqend{,} \nonumber \\
\label{eq:tildeT}
\eeqa
we find
\beq
Q_A^{(\bar{c})} = \int_\Sigma d\Sigma\,n_\mu \left[
\xi_{A\nu}\widetilde{\mathcal{T}}^{\mu\nu} - (\nabla_\alpha \xi_{A\nu})\bar{c}^\alpha 
H^{\mu\nu}\right]\eqend{.}
\label{eq:charge-cbar-new}
\eeq

The BRST transforms of the charges $Q_A^{(c)}=\delta_{\textrm{B}} Q_A^{(H)}$ vanish because $\delta_{\textrm{B}}^2 = 0$.  However, the BRST transforms of the charges $Q_A^{(\bar{c})}$ are nonzero. The conservation of $Q_A^{(\bar{c})}$ and the BRST invariance of the theory imply that the charges $\delta_{\textrm{B}} Q_A^{(\bar{c})}$ are also conserved.  We show in Appendix~\ref{AppB} that these charges are precisely the Noether charges $Q_A^{(\textrm{st})}$ associated with the spacetime symmetries generated by the Killing vectors $\xi_{A}^\mu$.

We note in passing that the action is invariant under the ``anti-BRST'' transformation if $k=1/2$ (the de~Donder gauge). (The anti-BRST 
invariance of Yang-Mills theory has been found in Refs.~\cite{Curci:1976bt,Ojima:1980da,Baulieu:1981sb}. The anti-BRST invariance has also been studied in some formulations of the gauge sector of general relativity, which appear different from ours~\cite{Nakanishi:1980rf,Delbourgo:1981sp,Hirayama:1982eb,Kuusk:1983qc}.) When $k = 1/2$, the last term in Eq.~(\ref{newlagrangian}) vanishes, and the Lagrangian density $\mathcal{L}_{\textrm{FP}}+\mathcal{L}_{\textrm{gf}}$ is left unchanged if we replace $(c^\mu,\bar{c}^\mu, \bar{B}^\mu)$ by $(\bar{c}^\mu, -c^\mu, B^\mu)$.  Hence, for this value of the gauge parameter the action is also invariant under the following anti-BRST transformation:
\begin{subequations} 
\beqa
\delta_{\bar{\textrm{B}}} h_{\mu\nu} & = & 
\nabla_\mu \bar{c}_\nu + \nabla_\nu \bar{c}_\mu + \pounds_{\bar{c}} 
h_{\mu\nu}\eqend{,}\label{anti-BRSTtr1}\\
\delta_{\bar{\textrm{B}}}\bar{c}^\mu & =  & \bar{c}^\alpha\nabla_\alpha \bar{c}^\mu\eqend{,}\\
\delta_{\bar{\textrm{B}}}c^\mu & = & - i \bar{B}^\mu\eqend{,}\\
\delta_{\bar{\textrm{B}}}\bar{B}^\mu & = & 0\eqend{.}\label{anti-BRSTtr4}
\eeqa
\end{subequations}
Furthermore, Eq.~\eqref{eq:Tmunu} shows that for $k = 1/2$ the tensor $\mathcal{T}_{\mu\nu}$ corresponds to the tensor $\mathcal{S}_{\mu\nu}$ given in Eq.~\eqref{mathcalC} with $c^\mu$ replaced by $\bar{c}^\mu$. This implies that $Q_A^{(\bar{c})} = \delta_{\bar{\textrm{B}}}Q_A^{(H)}$.  Hence, we conclude that the Noether charges associated with the background spacetime symmetries are the anti-BRST transforms of the charges $Q_A^{(c)}$.  Thus, the de~Donder gauge appears to be a natural one because all conserved charges found in this section can be derived from $Q^{(H)}_A$ by the BRST and anti-BRST transformation in this gauge.

\section{Identification of the Killing vector modes}                                                                                                %
\label{sec:identify-Killing}                                                                                                                        %

Since the charges $Q_A^{(H)}$, $Q_A^{(c)}$ and $Q_A^{(\bar{c})}$ found in the previous section are conserved, it is consistent to require that all physical state, particularly the vacuum state $|\Omega\rangle$, be annihilated by these charges:
\beq
Q_A^{(H)}|\Omega\rangle = Q_A^{(c)}|\Omega\rangle = Q_A^{(\bar{c})}
|\Omega\rangle = 0\eqend{.} \label{eq:the-conditions}
\eeq
The main aim of this paper is to show that imposing these conditions on $|\Omega\rangle$ corresponds to  using the FP-ghost propagator regularized by a finite mass term and then taking the massless limit at the end as described in Sec.~\ref{sec:FPghostdivergence}.  We shall discuss this equivalence in the Hamiltonian formulation in Sec.~\ref{sec:Hamiltonian}.   For this purpose we need to identify the components of the fields $B^\mu$, $c^\mu$ and $\bar{c}^\mu$ that are proportional to the Killing vectors at each time. The conserved charges $Q^{(H)}_A$, $Q^{(\bar{c})}_A$ and $Q^{(c)}_A$ 
will be shown to be (essentially) the canonical conjugate momenta of these components. This procedure may look rather artificial, but it is necessary for using the Hamiltonian formalism to discuss the conditions \eqref{eq:the-conditions}.

We first extract the modes proportional to the Killing vectors $\xi_{A}^\mu$ at each time $t$ for $V^\mu = c^\mu$, $\bar{c}^\mu$ and $B^\mu$ as
\beq\label{eq:zero_mode_projection}
V^{A}_{(0)}(t) = \int_\Sigma d\Sigma V^\mu \eta_\mu^{A}\eqend{,}
\eeq
where $\Sigma$ is the hypersurface of constant $t$, which is an $(n-1)$-dimensional sphere of radius $\cosh t$. The covectors $\eta_{\mu}^{A}$ are chosen to satisfy
\beq
\int_\Sigma d\Sigma\, \xi_{A}^\mu \eta_\mu^{B} = \delta_A^B\eqend{.}  \label{eq:integral-condition}
\eeq
These conditions do not determine $\eta^A_\mu$ uniquely and there is some freedom in choosing them. It is natural to choose them for the rotation Killing vectors as
\begin{equations}[eq:rotation-covectors]
\cosh^{n-1} t\,\eta_{(\sigma,\rR)0} & = 0\eqend{,}\\
\cosh^{n-1} t\,\eta_{(\sigma,\rR)i} & = Y_{(1\sigma)i}\eqend{.}
\end{equations}
As for the covectors associated with the boost Killing vectors, a simple choice is
\begin{equations}[eq:boost-covectors]
\cosh^{n-1}t\, \eta_{(\sigma,\rB)0} & = Y_{(1\sigma)}\eqend{,}\\
\cosh^{n-1}t\, \eta_{(\sigma,\rB)i} & =  0\eqend{.}
\end{equations}
With the help of these definitions, we can then expand the field $V^\mu$ as
\beq
V^\mu(x) = \sum_{A} V^{A}_{(0)}(t)\xi^\mu_{A}(x) + V^{\mu}_{(+)}(x)\eqend{.} \label{eq:V-decomposition}
\eeq
In order to simplify the notation, we also define 
\begin{equations}[eq:ghosts_zero_modes]
 \theta^A(t) & \equiv c^A_{(0)}(t)\eqend{,}\\
 \bar{\theta}^A(t) & \equiv \cbar^A_{(0)}(t)\eqend{.}
\end{equations}

In the ADM Hamiltonian formalism~\cite{Arnowitt:1962hi} the metric is given as follows:
\beq
ds^2 = - N^2 dt^2 + g_{ij}(dx^i + N^i dt)(dx^j + N^j dt)\eqend{,}
\eeq
where $N$ and $N^i$ are called the lapse function and shift vector, respectively. They are given in terms of the full metric components $\tilde{g}_{\mu\nu}$ as
\begin{subequations}
\beqa
N & = & \sqrt{-\tilde{g}_{00} + \tilde{g}^{ij}\tilde{g}_{0i}\tilde{g}_{0j}}\eqend{,}\\
N^i & = & \tilde{g}^{ij}\tilde{g}_{0j}\eqend{,}
\eeqa
\end{subequations}
where $\tilde{g}^{ij}$ is the inverse of the matrix $\tilde{g}_{ij}$.  As is well known, the Lagrangian density for the Einstein-Hilbert action can be given in terms of $\tilde{g}_{ij}$, $N$ and $N^i$ up to a total divergence, and this Lagrangian density contains no time derivatives of $N$ or $N^i$. (See, e.g.\ Appendix E of Ref.~\cite{wald}.)  

Since this Lagrangian density depends on $h_{00} = \tilde{g}_{00} -g_{00}$ and $h_{0i} = \tilde{g}_{0i} - g_{0i}$ only through $N$ and $N^i$,  it 
does not contain any time derivatives of $h_{00}$ or $h_{0i}$.  This allows us to identify $H_{0\nu}$ as the momentum variables conjugate to $B^\nu$ as
\beq
\sqrt{-g} H_{0\nu} = \frac{\partial \mathcal{L}_{\textrm{gf}}}{\partial \dot{B}^\nu}\eqend{,}
\eeq
with the notation $\dot{f} \equiv \del_t f$, if there were no terms containing $\dot{h}_{0\nu}$ in the FP-ghost Lagrangian density $\mathcal{L}_{\textrm{FP}}$. In fact, the Lagrangian density $\mathcal{L}_{\textrm{FP}}$ does contain terms involving $\dot{h}_{0\nu}$, but they can be removed by redefining the auxiliary field $B^\mu$.  Thus, by defining
\beq
\check{B}^\mu \equiv B^\mu - i (\nabla_\alpha\bar{c}^\mu)c^\alpha\eqend{,} \label{eq:checkB}
\eeq
we find, up to a total divergence,
\begin{splitequation}
& \mathcal{L}_{\textrm{FP}} + \mathcal{L}_{\textrm{gf}}\\
&\; =  - \nabla^\mu \check{B}^\nu H_{\mu\nu}  +  i  {{R^\nu}_{\beta \alpha}}^\mu \bar{c}^\beta c^\alpha H_{\mu\nu} \\
&\; \phantom{=}\;  - i \nabla^\mu \cbar^\nu\left(\nabla_\mu c_\nu + \nabla_\nu c_\mu - 2k g_{\mu\nu}\nabla_\alpha c^\alpha\right)(1+kh)\\
&\; \phantom{=}\;  - i \left[ - (\nabla^\mu \bar{c}^\nu)(\nabla_\alpha c^\alpha) + (\nabla_\alpha \cbar^\mu)(\nabla^\nu c^\alpha)\right.\\
&\; \phantom{=}\;\;\;\;\;\;\; \left.  + (\nabla^\alpha \cbar^\mu)(\nabla_\alpha c^\nu) + (\nabla^\mu \cbar^\alpha)(\nabla_\alpha c^\nu)\right] H_{\mu\nu} 
\eqend{.} \label{eq:alternativeL-1}
\end{splitequation}
We present a derivation of this result in Appendix~\ref{App:hamiltonian-approach}.

To identify the conserved charges found in the previous section essentially as the canonical momenta conjugate to cyclic variables we need to redefine the auxiliary field further.  Hence, let us define
\beq
\tilde{B}^\mu \equiv \check{B}^\mu  + i \bar{\theta}^{A} (\nabla_\alpha \xi_{A}^\mu) c^\alpha
+ i \cbar^{\alpha}_{(+)}(\nabla_\alpha \xi_{A}^\mu)\theta^{A}\eqend{,}
\eeq
where $\theta^A$ and $\bar{\theta}^A$ are the canonical variables multiplying the Killing vectors in the expansion of $c^\mu$ and $\bar{c}^\mu$, respectively, as defined in Eq.~(\ref{eq:ghosts_zero_modes}). Then, after a tedious but straightforward calculation we find
\begin{splitequation}
& \mathcal{L}_{\textrm{FP}} + \mathcal{L}_{\textrm{gf}}\\ 
&\; = \mathcal{L}_{\textrm{FP}+\textrm{gf}}^{(+)} + i \dot{\bar{\theta}}^A \dot{\theta}^B
\bigg\{ \left[ g_{ij} \xi_A^i \xi_B^j + \frac{2}{\beta} \xi_A^0 \xi_B^0\right] (1+kh) \\
&\; \phantom{=}  + (1-2k)\xi_A^0 \xi_B^\nu H_{0\nu} + \xi_A^\mu \xi_B^\nu H_{\mu\nu}\bigg\} + i \dot{\bar{\theta}}^A \theta^B \xi_{\left[A,B\right]}^\nu H_{0\nu} \\
&\; \phantom{=} + i \dot{\bar{\theta}}^A \left[ \xi_A^\nu \widetilde{\mathcal{S}}_{0\nu}^{(+)} - (\nabla_\alpha \xi_A^\nu)
c^\alpha_{(+)} H_{0\nu}\right] \\
&\; \phantom{=} + i\left[ \xi_A^\nu \widetilde{\mathcal{T}}_{0\nu}^{(+)} - (\nabla_\alpha \xi_A^\nu)\cbar^\alpha_{(+)} H_{0\nu}\right]\dot{\theta}^A + \dot{B}_{(0)}^A\xi^\nu_A H_{0\nu}\eqend{.}
\label{eq:theta-thetabar}
\end{splitequation}
In Eq.~(\ref{eq:theta-thetabar}), we have defined the Killing vector $\xi^\mu_{[A,B]} \equiv [\xi_A,\xi_B]^\mu$, the variable $B_{(0)}^A$ is the coefficient of the Killing vector mode of the field $B^\mu$ as defined by
Eq.~\eqref{eq:V-decomposition}, and $\mathcal{L}^{(+)}_{\textrm{FP}+\textrm{gf}
}$ does not contain variables
$B_{(0)}^A$, $\theta^A$ or $\bar{\theta}^A$.  The tensors $\widetilde{\mathcal{S}}^{(+)}_{\mu\nu}$ and
$\widetilde{\mathcal{T}}^{(+)}_{\mu\nu}$ are obtained by replacing $c^\mu$ and $\bar{c}^\mu$ by
$c_{(+)}^\mu$ and $\bar{c}_{(+)}^\mu$, which are defined by Eq.~\eqref{eq:V-decomposition}, 
in $\widetilde{\mathcal{S}}_{\mu\nu}$ in Eq.~\eqref{eq:tildeS}
and $\widetilde{\mathcal{T}}_{\mu\nu}$ in Eq.~\eqref{eq:tildeT}, respectively.  Then we find
\begin{subequations}
\beqa
\frac{\partial\ }{\partial \dot{B}_{(0)}^A}
(\mathcal{L}_{\textrm{FP}} + \mathcal{L}_{\textrm{gf}})
& = & \xi_A^\nu H_{0\nu}\eqend{,}\\
\frac{\partial\ }{\partial \dot{\bar{\theta}}^A}
(\mathcal{L}_{\textrm{FP}} + \mathcal{L}_{\textrm{gf}})
& = & i \left[\xi_A^\nu \widetilde{\mathcal{S}}_{0\nu} - (\nabla_\alpha \xi_{A}^\nu) c^\alpha H_{0\nu}\right]
\eqend{,}\nonumber \\ \\
\frac{\partial\ }{\partial \dot{\theta}^B}
(\mathcal{L}_{\textrm{FP}} + \mathcal{L}_{\textrm{gf}}) 
& = &  - i \left[\xi_B^\nu \widetilde{\mathcal{T}}_{0\nu} - (\nabla_\alpha \xi_{B}^\nu) 
\cbar^\alpha H_{0\nu}\right]
\nonumber \\
&& - i \bar{\theta}^A \xi^\nu_{\left[A,B\right]} H_{\nu 0}\eqend{,}
\eeqa
\end{subequations}
where the derivatives with respect to $\dot{\bar{\theta}}^A$ and $\dot{\theta}^B$ are left-derivatives. The integral of these over a Cauchy surface of constant time gives the canonical momenta conjugate to $B^A_{(0)}$, $\bar{\theta}^A$ and $\theta^B$, which will be denoted by $p_A$, $\varphi_A$ and $\bar{\varphi}_B$, respectively.  They satisfy
\beq
\big[p_A,B^B_{(0)}\big] = \left\{ \varphi_A,\bar{\theta}^B\right\} = \left\{ \bar{\varphi}_A,\theta^B\right\}
= - i\delta_A^B\eqend{,}
\eeq
where $\{ \omega_1,\omega_2\} \equiv \omega_1\omega_2 + \omega_2\omega_1$. Equations~\eqref{eq:bosonic-charge}, \eqref{eq:charge-c} and \eqref{eq:charge-cbar-new} then yield
\begin{subequations}
\beqa
p_A & = & Q_A^{(H)}\eqend{,}\label{pAandQA}\\
\varphi_A & = & iQ_A^{(c)}\eqend{,} \label{phiAandQA}\\
\bar{\varphi}_A & = & -i\left(Q_A^{(\bar{c})} +  \bar{\theta}^BQ_{\left[B,A\right]}^{(H)}\right)\eqend{,}
\label{eq:over-varphi}
\eeqa
\end{subequations}
where the charge $Q_{\left[B,A\right]}^{(H)}$ is the bosonic charge of Eq.~(\ref{eq:bosonic-charge}) corresponding to the Killing vector $\xi^\mu_{[B,A]}$. It is interesting to note that 
\beq
\left\{Q_A^{(\bar{c})},Q_B^{(c)}\right\} = -iQ_{[A,B]}^{(H)}\eqend{.}
\eeq
By applying the BRST transformation and using  $\delta_{\textrm{B}}Q_A^{(\cbar)} = 0$, $Q_A^{(\bar{c})} = \delta_{\bar{\textrm{B}}}Q_A^{(H)}$ and
$Q_A^{(\textrm{st})} = i \delta_{\bar{\textrm{B}}}Q_A^{(c)}$, we find
\beq
\left[ Q_A^{(\textrm{st})}, Q_B^{(c)}\right] =  Q_{[A,B]}^{(c)}\eqend{,}
\eeq
which is the expected action of the spacetime-symmetry charges $Q_A^{(\textrm{st})}$ on $Q_B^{(c)}$.

The canonical conjugate momenta $p_A$ and $\varphi_A$ are those of cyclic variables $B_{(0)}^A$ and $\bar{\theta}^A$ as can be seen from Eq.~\eqref{eq:theta-thetabar}, and they are indeed time independent, being proportional to conserved charges.  The time derivative of $\bar{\varphi}_A$ can be found from the Lagrangian density \eqref{eq:theta-thetabar} as
\beqa
\dot{\bar{\varphi}}_A & = & \frac{\partial\ }{\partial\theta_A}
\int_{\Sigma} d\Sigma\, (\mathcal{L}_{\textrm{FP}} + \mathcal{L}_{\textrm{gf}}) \nonumber \\
& = & - i\dot{\bar{\theta}}^B Q_{\left[B,A\right]}^{(H)}\eqend{,}
\eeqa
which agrees with the result obtained by differentiating Eq.~\eqref{eq:over-varphi} directly and using the conservation of the charges $Q_A^{(\bar{c})}$ and $Q^{(H)}_{\left[B,A\right]}$.

\section{The conditions on the vacuum state at tree level}                                                                                          %
\label{sec:Killing-vector-modes}                                                                                                                    %

As we stated before, the main purpose of this paper is to show that the use of the regularized FP-ghost propagator for perturbative gravity in de~Sitter spacetime corresponds to the conditions $Q^{(H)}_A|\Omega\rangle = Q^{(c)}_A|\Omega\rangle = Q^{(\cbar)}_A|\Omega\rangle = 0$ on the
vacuum state $|\Omega\rangle$.  In this section, we show that the use of the regularized FP-ghost propagator implies that the non-interactive vacuum state $|0\rangle$ is annihilated by the tree-level charges. From here to the end of this section, the charges $Q^{(H)}_A$, $Q^{(c)}_A$ and $Q^{(\cbar)}_A$ are the conserved charges in the noninteracting theory with the interactions turned off, which are linear in $h_{\mu\nu}$, $c^\mu$ and $\cbar^\mu$, respectively.

For the bosonic charges $Q^{(H)}_A$ we show in Appendix~\ref{non-contrib} that the result $Q^{(H)}_A|0\rangle = 0$ or, more precisely,  $\langle 0|\omega Q^{(H)}_A|0\rangle = 0$ for any canonical variable $\omega$ except $B_{(0)}^A$, which are canonically conjugate to $Q^{(H)}_A$, follows automatically in the standard de~Sitter-invariant quantization of linearized gravity in the Landau gauge.

\subsection{Scalar field zero mode}

To illustrate in what way the regularized propagator corresponds to the charges $Q^{(c)}_A$ and $Q^{(\cbar)}_A$ annihilating the vacuum state at linear level in the massless limit, let us consider a massive Hermitian scalar field $\phi$ on de~Sitter spacetime. Expanding this field operator in terms of the scalar spherical harmonics, Eq.~(\ref{eq:scalar_spherical_harmonics}), we obtain
\begin{equation}
 \phi(t,\boldsymbol{\theta}) = \sum_{\ell=0}^\infty\sum_\sigma\left[a_{\ell\sigma}f_\ell(t)Y_{(\ell\sigma)}(\boldsymbol{\theta}) + a^\dagger_{\ell\sigma}f_\ell^*(t)Y_{(\ell\sigma)}^*(\boldsymbol{\theta}) \right]\eqend{,}
\end{equation}
where $[a_{\ell\sigma},a^\dagger_{\ell'\sigma'}] = \delta_{\ell\ell'}\delta_{\sigma\sigma'}$, with other commutators null, and $f_\ell$ are normalized according to the Klein-Gordon inner product and chosen such that we have the Bunch-Davies vacuum. The field time evolution is dictated by the Hamiltonian operator
\begin{equation}\label{eq:scalar_hamiltonian}
 H = \frac{1}{2}\left[\frac{\pi_{0}(t)\pi_{0}(t)}{\cosh^{n - 1}t} + m^2\cosh^{n - 1}t\phi_0(t)\phi_0(t)\right] + H_{(+)}\eqend{,}
\end{equation}
where $\phi_0 \equiv a_{00}f_0 + a^\dagger_{00}f_0^*$, $\pi_0 \equiv \cosh^{n - 1}t\, d\phi_0/dt$, thus $[\phi_0,\pi_0] = i$, and $H_{(+)}$ is the Hamiltonian operator of the modes with $\ell > 0$. We focus on the $\ell = 0$ mode, as it is the one responsible for the IR divergence of the propagator in this example. The form of $f_0$ in the small-$m$ limit can be found in Ref.~\cite{Gibbons:2014zya} and reads
\begin{equation}
 f_0(t) = \sqrt{\frac{V_{S^{n - 1}}}{2b_0}}\left\{\frac{1}{m} - m[g(t) + b_1 + ib_0f(t)]\right\} + \mathcal{O}(m^2)\eqend{,}
\end{equation}
where $b_0$ and $b_1$ are constants, $ V_{S^{n - 1}} \equiv 2\pi^{n/2}/\Gamma\left(\frac{n}{2}\right)$ is the volume of the unit $S^{n - 1}$, and
\begin{equations}
 f(t) & \equiv \int_0^t\frac{dt'}{V(t')}\eqend{,}\label{eq:f}\\
 g(t) & \equiv \int_0^t\frac{dt'}{V(t')}\int_0^{t'}dt''V(t'')\label{eq:g}\eqend{,} 
\end{equations}
where we have defined $V(t) \equiv V_{S^{n - 1}}\cosh^{n - 1}t$. The contribution coming from the zero mode to the propagator and its time derivatives in the de~Sitter invariant vacuum $|0\rangle$ has the form
\begin{splitequation}\label{eq:scalar_zero_mode_correlator}
 \langle 0|\phi_0(t)\phi_0(t')|0\rangle 
 & = f_0(t)f_0^*(t')\\
 & = \frac{V_{S^{n - 1}}}{2b_0}\frac{1}{m^2} - \sqrt{\frac{V_{S^{n - 1}}}{2b_0}}\big\{g(t) + g(t')\\ 
 & \phantom{=} + 2b_1 + ib_0[f(t) - f(t')]\big\} + \mathcal{O}(m^2)\eqend{,}
\end{splitequation}
\begin{splitequation}
 \langle 0|\phi_0(t)\pi_0(t')|0\rangle 
 & = f_0(t)\cosh^{n - 1}t'\dot{f}_0^*(t')\\
 & = - \frac{V(t')[\dot{g}(t') - ib_0\dot{f}(t')]}{\sqrt{2V_{S^{n - 1}}b_0}} + \mathcal{O}(m^2)\eqend{,}
\end{splitequation}
and
\begin{equation}\label{eq:scalar_zero_mode_momentum_correlator}
 \langle 0|\pi_0(t)\pi_0(t')|0\rangle = \mathcal{O}(m^2)\eqend{.}
\end{equation}
Hence, in the massless limit one finds $\langle 0|\pi_0(t)\pi_0(t')|0\rangle = 0$. This can be stated as $\langle 0|\omega(t)\pi_0(t')|0\rangle = 0$, where $\omega$ is any canonical variable except $\phi_0$.

If one expects the field $\phi$ to represent an observable or to couple with another field through its amplitude, then Eqs.~(\ref{eq:scalar_zero_mode_correlator}) -~(\ref{eq:scalar_zero_mode_momentum_correlator}) are just a manifestation of the well-known fact that there is no de~Sitter-invariant state for a massless scalar field~\cite{Allen:1985ux,Allen:1987tz}.\footnote{Interestingly enough, in the Euclidean theory it is possible to show that a massless free scalar field on $S^n$ admits fully symmetric states~\cite{Folacci:1992xc}. The idea is to treat the massless field as a gauge theory symmetric under $\phi \to \phi + \textrm{constant}$ and add a gauge-fixing term to the Lagrangian that effectively removes the zero mode.} On the other hand, if the field $\phi(t)$ itself is unobservable and only interacts via its derivatives, which is precisely the case of the FP ghosts, then the correlators $\langle 0|\phi_0(t)\phi_0(t')|0\rangle \to \infty$ and $\langle 0|\phi_0(t)\pi_0(t')|0\rangle \neq 0$ in the limit $m \to 0$ are irrelevant for observable quantities. Moreover, it is clear from the form of the Hamiltonian~(\ref{eq:scalar_hamiltonian}) that $\phi_0$ is a cyclic variable in the massless limit. Its canonical conjugate momentum $\pi_0$ is the conserved charge associated with the conserved current $J_\mu \equiv \nabla_\mu\phi$. This allows us to interpret the statement $\langle 0|\omega(t)\pi_0(t')|0\rangle = 0$ simply as
\begin{equation}\label{eq:scalar_charge_condition}
 \pi_0|0\rangle = 0\eqend{.}
\end{equation}
The condition~(\ref{eq:scalar_charge_condition}) is the requirement that the state $|0\rangle$ is invariant under $\phi \to \phi + \textrm{constant}$, which is a gauge transformation for the massless scalar field.

We can turn the argument above around and show that the condition~(\ref{eq:scalar_charge_condition}) in the massless theory corresponds to discarding the contribution of the zero mode to the propagator. From the Hamiltonian~(\ref{eq:scalar_hamiltonian}) and the Heisenberg equation, we have that
\begin{equation}
 \frac{d\ }{dt}\left[V(t)\frac{d\phi_0}{dt}\right] = 0\eqend{.}
\end{equation}
Therefore, the zero mode $\phi_0$ is analogous to a free quantum particle and we can expand the field operator $\phi$ as
\begin{splitequation}
 \phi(t,\boldsymbol{\theta}) 
 & = \hat{q} + \hat{p}f(t)\\ 
 & \phantom{=} + \sum_{\ell=1}^\infty\sum_\sigma\left[a_{\ell\sigma}f_\ell(t)Y_{(\ell\sigma)}(\boldsymbol{\theta}) + a^\dagger_{\ell\sigma}f_\ell^*(t)Y_{(\ell\sigma)}^*(\boldsymbol{\theta}) \right]\eqend{,}
\end{splitequation}
where $f(t)$ was defined in Eq.~(\ref{eq:f}) and $[\hat{q},\hat{p}] = i$. We note that $\phi_0(0) = \sqrt{V_{S^{n - }}}\hat{q}$ and $\pi_0 = \sqrt{V_{S^{n - }}}\hat{p}$. As in quantum mechanics, we can represent the operators $\hat{q}$ and $\hat{p}$ on $L^2(\mathbb{R})$ as the multiplication by $q$ and the derivative $-id/dq$, respectively. We then consider the field state $|\Psi\rangle = \psi(q)\otimes |0_{(+)}\rangle$, where $\psi(q)$ is a normalized wave function and $|0_{(+)}\rangle$ is the vacuum state for the modes with $\ell > 0$, i.e.\ $a_{\ell\sigma}|0_{(+)}\rangle = 0$ for all $\ell \geq 1$. It is possible to show (see, e.g.\ Refs.~\cite{Kirsten:1993ug, Gibbons:2014zya}) that the state $|\Psi\rangle$ is de~Sitter invariant if, and only if, condition~(\ref{eq:scalar_charge_condition}) is satisfied, i.e.\ $\hat{p}|\Psi\rangle = 0$. Hence, for $|\Psi\rangle$ to be de~Sitter invariant we must have the wave function $\psi(q)$ constant. Again, if $\phi$ is observable or couples through its amplitude, this is a manifestation of the fact that no such $|\Psi\rangle$ exists, since $\int_{-\infty}^{+\infty}dq|\psi(q)|^2 = \infty$. However, if our field is unobservable and interacts via its derivatives, then, since the zero mode is spatially constant and its time derivative annihilates the state, we can simply ignore it by redefining the inner product of the field space of states. Thus, for two field states $|\Psi_1\rangle = \psi_1(q)\otimes |\alpha_{(+)1}\rangle$ and $|\Psi_2\rangle = \psi_2(q)\otimes |\alpha_{(+)2}\rangle$, where $|\alpha_{(+)1}\rangle$ and $|\alpha_{(+)2}\rangle$ are  states  in  the  Fock  space  built by applying $a_{\ell\sigma}^\dagger$ on $|0_{(+)}\rangle$, we define $\langle\Psi_1|\Psi_2\rangle \equiv \langle\alpha_{(+)1}|\alpha_{(+)2}\rangle$.\footnote{An alternative construction for the Fock space can be found in Refs.~\cite{Bertola2007, Epstein:2014jaa}, which mirrors the Gupta-Bleuler quantization method for the electromagnetic field.} The result of computing the propagator in the vacuum state annihilated by $\hat{p}$, with the redefined inner product, is the scalar counterpart of the use of the regularized propagator discussed in Sec.~\ref{sec:FPghostdivergence}.

\subsection{Ghost fields zero modes in perturbative quantum gravity}

Let us now return to the analysis of the FP-ghost propagator in perturbative quantum gravity. What we shall demonstrate at tree level is that, if we regularize the propagator with a small mass and then take the massless limit,  then $\langle 0| Q^{(c)}_A(t)\Lambda|0\rangle = 0$ and $\langle 0|\omega Q^{(\cbar)}_A(t)|0\rangle = 0$ for any canonical variables $\omega$ unless $\omega= \bar{\theta}^A$ for the former and unless $\omega = \theta^A$ for the latter.  (Note that $\bar{\theta}^A$ and $\theta^A$ are canonically conjugate to $Q^{(c)}_A$ and $Q^{(\cbar)}_A$, respectively, at tree level.)  It is sufficient to show that $\langle 0|Q^{(c)}_A(t) Q^{(\bar{c})}_B(t')|0\rangle$, $\langle 0| Q^{(c)}_A(t)\bar{c}_{(+)}^\mu(x')|0\rangle$ and $\langle 0|c_{(+)}^\mu(x) Q^{(\bar{c})}_B(t')|0\rangle$ all vanish in the massless limit.  Some details for the following discussion will be delegated to Appendix~\ref{app-FPscalar-no-contrib}.

The FP-ghost field equation with mass $m$ at tree level reads
\beq
\nabla_\nu (\nabla^\nu c^\mu + \nabla^\mu c^\nu - k g^{\mu\nu}\nabla_\alpha c^\alpha) - m^2 c^\mu = 0
\eqend{.}
\eeq
There are two types of solutions to this equation.  By writing
\beq
c^\mu = V^\mu + \nabla^\mu\Phi\eqend{,}
\eeq
where $\nabla_\mu V^\mu = 0$, we find
\begin{subequations}
\beqa
&& \Box\Phi - \frac{\beta}{2}\left[2(n-1)-m^2\right]\Phi  = 0\eqend{,} \label{eq:scalar-sector-ghost}\\
&& \Box V_\mu - \left[ m^2 - (n-1)\right]V_\mu = 0\eqend{,}
\eeqa
\end{subequations}
where we have defined $\Box \equiv \nabla^\mu\nabla_\mu$. The scalar sector $\nabla_\mu \Phi$ contributes to the charge $Q_A^{(c)}$ at tree level only at order $m^2$, as shown in Appendix~\ref{app-FPscalar-no-contrib}. Hence, we do not need to consider this sector in calculating $\langle 0|\omega Q_A^{(\bar{c})}|0\rangle$ and $\langle 0|Q_A^{(c)}\bar{\omega}|0\rangle$ at tree level in the $m\to 0$ limit. For positive $\beta$, the field $\Phi$ is a scalar field of positive mass, and there is no divergence from this sector in the limit $m\to 0$. For $\beta$ negative, but not in the set $-s(s + n - 1)/(n - 1)$, with $s = 0, 1,2,\dots$, there is also no divergence for the scalar sector, even though the scalar propagator grows as the spacetime points become largely separated. We will return to this point in Sec.~\ref{sec:summary}.

The mode functions constituting the vector sector, $V_\mu$, are given by 
\begin{equations}  \label{eq:V_1_mode}
 V_0^{(1;\ell,\sigma)} & = 0\eqend{,}\\
 V_i^{(1;\ell,\sigma)} & = \frac{C_m^\ell}
{\cosh^\frac{n - 4}{2} t}\textrm{P}^{-\mu_\ell}_{-\frac{1}{2} + \lambda_2}(i\sinh t)Y_{(\ell\sigma)i}
\eqend{,}
\end{equations}
and
\begin{equations} \label{eq:V_2_mode}
 V_0^{(2;\ell,\sigma)} 
& = - \sqrt{\frac{\ell(\ell+n-2)}{2(n-1)-m^2}} C_m^\ell \nonumber \\
&\phantom{=} \times \frac{1}{\cosh^\frac{n}{2} t}\textrm{P}^{-\mu_\ell}_{-\frac{1}{2} + \lambda_2}(i\sinh t)Y_{(\ell\sigma)}
\eqend{,}\\
 V_i^{(2;\ell,\sigma)} 
& = \frac{C_m^\ell}{\sqrt{\ell(\ell+n-2)\left[2(n-1)-m^2\right]}}\nonumber \\
&\phantom{=} \times \left[\cosh^2t\,\del_t + (n - 1)\sinh t\cosh t\right]\nonumber \\
&\phantom{=} \times \frac{1}{\cosh^{\frac{n}{2}}t} \textrm{P}^{-\mu_\ell}_{-\frac{1}{2} + \lambda_2}(i\sinh t)\mathcal{D}_iY_{(\ell\sigma)}\eqend{,}
\end{equations}
where
\begin{equations}
C_m^\ell & \equiv 
\sqrt{\frac{\Gamma\left(\ell + \frac{n - 1}{2} - \lambda_2\right)\Gamma
\left(\ell + \frac{n - 1}{2} + \lambda_2\right)}{2}}\eqend{,}\\
\lambda_2 & \equiv \sqrt{\left( \frac{n+1}{2}\right)^2 - m^2}\eqend{,}\\
\mu_\ell & \equiv \ell+\frac{n-2}{2}\eqend{.}
\end{equations}
The function $\mathrm{P}_\nu^{-\mu}(x)$ is the associated Legendre function of the first 
kind~\cite{gradshteyn2007}. Then, we use the covectors~(\ref{eq:rotation-covectors}) and~(\ref{eq:boost-covectors}) to extract the zero-mode part of the vector modes with $\ell = 1$, which yields
\begin{equation}\label{eq:V_1_1_mode}
V_\mu^{(1;1,\sigma)} = \frac{C^1_m}{\cosh^\frac{n}{2}t}\mathrm{P}^{-\frac{n}{2}}_{-\frac{1}{2} + \lambda_2}(i\sinh t)\xi_{(\sigma,\mathrm{R})\mu}
\end{equation}
and
\begin{splitequation}\label{eq:V_2_1_mode}
V_\mu^{(2;1,\sigma)} 
& = \sqrt{\frac{n - 1}{2(n - 1) - m^2}}\frac{C^1_m}{\cosh^\frac{n}{2}t}\mathrm{P}^{-\frac{n}{2}}_{-\frac{1}{2} + \lambda_2}(i\sinh t)\xi_{(\sigma,\mathrm{B})\mu}\\
&\phantom{=} + \frac{C^1_m}{\sqrt{(n - 1)[2(n - 1) - m^2]}}\\
&\phantom{=}\times \frac{d\ }{dt}\left[\frac{1}{\cosh^\frac{n}{2}t}\mathrm{P}^{-\frac{n}{2}}_{-\frac{1}{2} + \lambda_2}(i\sinh t)\right]\chi_{(\sigma)\mu}\eqend{,}
\end{splitequation}
where the Killing vectors were given in Eqs.~(\ref{eq:rotation-Killing}) and~(\ref{eq:boost-Killing}) and we have defined the vectors
\begin{equations}
\chi^0_{(\sigma)} & = 0\eqend{,}\\
\chi^i_{(\sigma)} & = \mathcal{D}^iY_{(1,\sigma)}\eqend{.}
\end{equations}

The vector sector of the FP-ghost field can be expanded as
\begin{equation}\label{eq:ghost_field_expansion}
 V_\mu = \sum_{I,\ell,\sigma}\left[
\alpha_{\ell\sigma}^{(I)}V_\mu^{(I;\ell,\sigma)} 
 + \alpha_{\ell\sigma}^{(I)\dagger}V_\mu^{(I;\ell\sigma)*}\right]\eqend{.} 
\end{equation}
The vector sector of the antighost field is expanded in the same way with the annihilation and creation operators, $\alpha_{\ell\sigma}^{(I)}$ and
$\alpha_{\ell\sigma}^{(I)\dagger}$, replaced by  $\bar{\alpha}_{\ell\sigma}^{(I)}$ and $\bar{\alpha}_{\ell\sigma}^{(I)\dagger}$, respectively.  The mode functions $V_\mu^{(I;\ell,\sigma)}$ are normalized so that these annihilation and creation operators satisfy
\begin{eqnarray}
 \left\{\alpha_{\ell\sigma}^{(I)}, \bar{\alpha}_{\ell'\sigma'}^{(J)\dagger}\right\} & = & 
 (-1)^{I+1} i\delta_{\ell\,\ell'}\delta_{\sigma\,\sigma'}\delta^{IJ}\eqend{,} \label{eq:anticomm-rel}
\end{eqnarray}
with all other anticommutators among $\alpha_{\ell\sigma}^{(I)}$ and $\bar{\alpha}_{\ell\sigma}^{(I)}$ and their Hermitian conjugates vanishing.

The de~Sitter invariant tree-level vacuum state $|0\rangle$ is annihilated by the annihilation operators $\alpha_{\ell\sigma}^{(I)}$ and $\bar{\alpha}_{\ell\sigma}^{(I)}$, $I=1,2$. It is useful to note that~\cite{gradshteyn2007}
\begin{eqnarray}
 \textrm{P}^{-\mu_\ell}_{-\frac{1}{2} + \lambda_2}(i\sinh t) 
& =&  \frac{(\cosh t)^{\mu_\ell}}{2^{\mu_\ell}\Gamma\left(\ell + 
\frac{n}{2}\right)} \nonumber \\
&& \times \phantom{}_2F_1\left(b^+_\ell, b^-_\ell; \ell + \frac{n}{2}; \frac{1 - i\sinh t}{2}\right)\eqend{,} \nonumber \\
\label{eq:hypergeometric}
\end{eqnarray}
where we have defined
\begin{equation}
 b^\pm_\ell \equiv \ell + \frac{n - 1}{2} \pm \lambda_2\eqend{.}
\end{equation}
The function $\phantom{}_2F_1(a,b;c;z)$ denotes Gauss's hypergeometric function.

The contribution to the conserved charges $Q^{(c)}_A$ and $Q^{(\cbar)}_A$ comes from the modes with $\ell=1$.  For $\ell=1$, the massless limit yields
\beq
\lim_{m\to 0}\mathrm{P}_{-\frac{1}{2}+\lambda_2}^{-\mu_1}(i\sinh t) = \frac{(\cosh t)^{\frac{n}{2}}}
{2^{\frac{n}{2}}\Gamma(\frac{n+2}{2})}\eqend{,}
\label{eq:m0-limit1}
\eeq
since $b_1^{-}\to 0$ as  $m\to 0$ and $\phantom{}_2F_1(\alpha,0;\gamma;z) = 1$. Moreover, in this limit we also obtain
\begin{equations}[eq:m0-limit2]
C_m^{1} & \approx \sqrt{\frac{\Gamma(n+2)}{2m^2}}\eqend{,}\\
\frac{1}{\sqrt{2(n - 1) - m^2}} &\approx \frac{1}{\sqrt{2(n - 1)}}\left[1 + \frac{m^2}{4(n - 1)}\right]\eqend{.}
\end{equations}
By substituting Eqs.~\eqref{eq:m0-limit1} and \eqref{eq:m0-limit2} into Eqs.~\eqref{eq:V_1_1_mode} 
and \eqref{eq:V_2_1_mode} we find that the leading terms for $\ell = 1$ are
\begin{equations}[eq:V1_V_2]
V_\mu^{(1;1,\sigma)} & \approx \frac{1}{\sqrt{2c_0}\,m}\xi_{(\sigma,\rR)\mu}\eqend{,} \label{eq:V1} \\
V_\mu^{(2;1,\sigma)} & \approx \frac{1}{2\sqrt{c_0}\,m}\xi_{(\sigma,\rB)\mu}\eqend{.} \label{eq:V2}
\end{equations}
We have used the doubling formula for the $\Gamma$-function to arrive at Eq.~\eqref{eq:V1_V_2}. The constant $c_0$, whose exact value is not important, is given by Eq.~\eqref{eq:c0-def}.

As we have stated before, only the $\ell=1$ modes contribute to the conserved FP-ghost charges at tree level.  By substituting the mode functions $V_\mu^{(1;1,\sigma)}$ and $V_\mu^{(2;1,\sigma)}$ given by Eqs.~\eqref{eq:V_1_1_mode} and~(\ref{eq:V_2_1_mode}) in Eq.~(\ref{eq:charge-c}) at tree level, and using Eq.~(\ref{eq:m0-limit2}), we find
\begin{splitequation}\label{eq:ghost_charges_hypergeometric_derivative_rotation}
& Q_{(\sigma,\rR)}^{(c)}(t)\\
& \approx  \frac{\cosh^{n+1}t}{\sqrt{2c_0}\,m}\frac{d\ }{dt}\phantom{}_2F_1\left(b_1^+,b_1^-; \frac{n+2}{2};\frac{1-i\sinh t}{2}\right)
\alpha_{1\sigma}^{(1)}\\ 
&\phantom{\approx}+ \textrm{H.c.} \eqend{,}
\end{splitequation}
\begin{splitequation}\label{eq:ghost_charges_hypergeometric_derivative_boost}
& Q_{(\sigma,\rB)}^{(c)}(t) \\
& \approx - \frac{\cosh^{n - 1}t}{\sqrt{c_0}\, m}\left(1 - \frac{n - 1}{2}\sinh^2 t - \frac{1}{2}\sinh t\cosh t\frac{d\ }{dt}\right)\\
&\phantom{\approx}\times \frac{d\ }{dt}\phantom{}_2F_1\left(b_1^+,b_1^-; \frac{n+2}{2};\frac{1-i\sinh t}{2}\right)\alpha_{1\sigma}^{(2)}+ \textrm{H.c.} \eqend{,}
\end{splitequation}
where H.c.\ stands for the Hermitian conjugate of the preceding terms.

The derivative of the hypergeometric function appearing above is evaluated in the small-$m$ limit in Appendix~\ref{app-FPscalar-no-contrib}, and it yields
\beqa
&& \frac{d\ }{dt}\phantom{}_2F_1\left(b_1^+,b_1^-; \frac{n+2}{2};\frac{1-i\sinh t}{2}\right) \nonumber \\
&& \approx - \frac{m^2}{\cosh^{n+1}t} \left( ic_0 + \int_0^t \cosh^{n+1}\tau\,d\tau\right)\eqend{.}
\label{eq:interesting}
\eeqa
Hence, substituting this result in Eqs.~(\ref{eq:ghost_charges_hypergeometric_derivative_rotation}) and~(\ref{eq:ghost_charges_hypergeometric_derivative_boost}) yields
\begin{splitequation}
& Q_{(\sigma,\mathrm{R})}^{(c)}(t)\\
& \approx  - \frac{m}{\sqrt{2c_0}}  \left( ic_0 + \int_0^t \cosh^{n+1}\tau\,d\tau\right)
\alpha^{(1)}_{1\sigma} + \textrm{H.c.}
\end{splitequation}
and
\begin{splitequation}
& Q_{(\sigma,\rB)}^{(c)}(t)\\ 
& \approx \frac{m\cosh^{n - 1}t}{\sqrt{c_0}}\left(1 - \frac{n - 1}{2}\sinh^2 t - \frac{1}{2}\sinh t\cosh t\frac{d\ }{dt}\right)\\
&\phantom{\approx}\times\frac{1}{\cosh^{n+1}t} \left( ic_0 + \int_0^t \cosh^{n+1}\tau\,d\tau\right)\alpha_{1\sigma}^{(2)}+ \textrm{H.c.} \eqend{.}
\end{splitequation}
By combining these equations with Eq.~\eqref{eq:V1_V_2} and the anticommutators~\eqref{eq:anticomm-rel}, we find for the rotation Killing vectors
\begin{splitequation}
& \langle 0| Q_{(\sigma,\rR)}^{(c)}(t)\bar{c}^\mu(x')|0\rangle\\
&\, = \frac{1}{2}\left( 1 - i c_0^{-1}\int_0^t \cosh^{n+1}\tau\,d\tau\right)\xi_{(\sigma,\rR)}^\mu(x')\eqend{,}
\label{eq:QRc-cbar}
\end{splitequation}
while for the boost Killing vectors we obtain 
\begin{splitequation}
& \langle 0| Q_{(\sigma,\rB)}^{(c)}(t)\bar{c}^\mu(x')|0\rangle\\
& = \frac{1}{2}\cosh^{n - 1}t\left(1 - \frac{n - 1}{2}\sinh^2 t - \frac{1}{2}\sinh t\cosh t\frac{d\ }{dt}\right)\\
&\phantom{=} \times\frac{1}{\cosh^{n + 1}t}\left( 1 - i c_0^{-1}\int_0^t \cosh^{n+1}\tau\,d\tau\right)\xi_{(\sigma,\rB)}^\mu(x')\eqend{.}
\label{eq:QBc-cbar}
\end{splitequation}
We similarly have
\begin{splitequation}
& \langle 0|c^\mu(x)Q_{(\sigma,\rR)}^{(\bar{c})}(t')|0\rangle\\
&\, = -\frac{1}{2}\left( 1 + i c_0^{-1}\int_0^{t'} \cosh^{n+1}\tau\,d\tau\right)\xi_{(\sigma,\rR)}^\mu(x)
\label{eq:QRcbar-c}
\end{splitequation}
and
\begin{splitequation}
& \langle 0|c^\mu(x)Q_{(\sigma,\rB)}^{(\bar{c})}(t')|0\rangle\\
& = -\frac{1}{2}\cosh^{n - 1}t'\left(1 - \frac{n - 1}{2}\sinh^2 t' - \frac{1}{2}\sinh t'\cosh t'\frac{d\ }{dt'}\right)\\
&\phantom{=} \times\frac{1}{\cosh^{n + 1}t'}\left( 1 + i c_0^{-1}\int_0^{t'} \cosh^{n+1}\tau\,d\tau\right)\xi_{(\sigma,\rB)}^\mu(x)\eqend{.}
\label{eq:QBcbar-c}
\end{splitequation}
These equations imply that
\begin{subequations}
\beqa
\langle 0|Q_A^{(c)}Q_B^{(\bar{c})}|0\rangle & = & 0\eqend{,} \\
\langle 0|Q_A^{(c)}\bar{c}_{(+)}^\mu(x)|0\rangle & = & 0\eqend{,}\\
\langle 0|c_{(+)}^\mu(x)Q_A^{(\bar{c})}|0\rangle & = & 0\eqend{,}
\eeqa
\end{subequations}
which can be summarized as
\beq
\langle 0|Q_A^{(c)}\bar{\omega}|0\rangle = \langle 0|\omega Q^{(\cbar)}_A|0\rangle = 0\eqend{,}
\eeq
for any canonical variables except for $\bar{\omega} =\bar{\theta}^A$ or  $\omega = \theta^A$.

\section{Hamiltonian perturbation theory}                                                                                                           %
\label{sec:Hamiltonian}                                                                                                                             %

In the previous section we showed that the small-mass regularization of the FP-ghost propagator corresponds to the vacuum state $|0\rangle$ being annihilated by the conserved charges $Q^{(c)}_A$ and $Q^{(\cbar)}_A$ at tree level in de~Sitter spacetime. The analogous result for the bosonic charge $Q^{(H)}_A$, i.e.\ that it annihilates the vacuum state at tree level in the standard de~Sitter-invariant quantization of linearized gravity in the Landau gauge, can be found in Appendix~\ref{non-contrib}. In this section we show that these charges annihilate the interacting vacuum state $|\Omega\rangle$ to all orders in Hamiltonian perturbation theory with $|\Omega\rangle$ defined in this framework.  That is, we show that the conditions $Q^{(X)}_A|0\rangle = 0$ for $X=H,c,\cbar$ at tree level are inherited in the interacting theory as $Q^{(X)}_A|\Omega\rangle = 0$.

Let us first elaborate on the meaning of the conditions $Q^{(X)}_A|\Omega\rangle = 0$ for $X=H,c,\cbar$. Since $Q^{(H)}_A = p_A$ are the canonical momenta conjugate to $B_{(0)}^A$, if $\Psi_\Omega(B_{(0)}^A,\cdots)$ is the Schr\"odinger representation of the state $|\Omega\rangle$, the operator $Q^{(H)}_A$ is represented by $- i \partial/\partial B_{(0)}^A$.  Hence, the condition $Q^{(H)}_A|\Omega\rangle = 0$ means that the corresponding Schr\"odinger wave function $\Psi_\Omega$ does not depend on the variables $B_{(0)}^A$. The charge $Q^{(H)}_A$ are the generators of the translation in the variables $B_{(0)}^A$. Hence, we may interpret the conditions $Q^{(H)}_A|\Omega\rangle = 0$ as the requirement that the vacuum state 
$|\Omega\rangle$ be invariant under the gauge transformation $B_{(0)}^A + \textrm{constant}$.  These are the natural conditions because the Hamiltonian is invariant under these gauge transformation, being independent of $B_{(0)}^A$.

Once the conditions $Q^{(H)}_A|\Omega\rangle = 0$ are imposed, we may set $p_A=Q^{(H)}_A = 0$ in the Hamiltonian for the purpose of evaluating the expectation values of operators not including $B_{(0)}^A$ in the vacuum state $|\Omega\rangle$. (We may exclude the variables $B_{(0)}^A$ since these are ``gauge-dependent variables'' breaking the gauge invariance generated by $Q^{(H)}_A$.) Then,  Eq.~\eqref{eq:theta-thetabar}, which shows that the only undifferentiated variables $\theta^A$ is multiplied by $Q^{(H)}_A$ in the Lagrangian, and  Eqs.~\eqref{phiAandQA} and \eqref{eq:over-varphi} imply that, after setting $Q^{(H)}_A = 0$, the charges $Q^{(c)}_A$ and $Q^{(\cbar)}_A$ are also effectively the canonical momenta conjugate to the variables $\bar{\theta}^A$ and $\theta^A$, respectively.  Hence, these fermionic conserved charges can also be regarded as generating the gauge transformation of adding constant Grassmann numbers to $\bar{\theta}^A$ and $\theta^A$.  

Since the conditions $Q^{(X)}_A|\Omega\rangle = 0$ with $X=H,c,\cbar$ enforces the gauge invariance of the Hamiltonian on the vacuum state, it is natural to expect that these condition at tree level, $Q^{(X)}_A|0\rangle = 0$, will lead to the same conditions after including the interaction.  
We propose a definition of the vacuum state in Hamiltonian perturbation theory for which this is indeed the case.

The interaction Hamiltonian density in theories with derivative interactions, such as perturbative gravity, is noncovariant. For this reason Hamiltonian perturbation theory is not widely used, unlike Lagrangian perturbation theory in the path-integral framework. The two perturbation schemes are equivalent in quantum electrodynamics (QED) with charged scalar field~\cite{Itzykson:1980rh}.  This equivalence is explained in Appendix~\ref{app-scalarQED}.  One can demonstrate the equivalence of the two schemes in a wide class of theories with derivative interactions including perturbative gravity~\footnote{A.~Higuchi and W.~C.~C.~Lima, in preparation.}.  

In Hamiltonian perturbation theory in Minkowski spacetime the expectation value of the time-ordered product $T \omega_1(t_1)\omega_2(t_2)\cdots\omega_N(t_N)$, where $\omega_1(t_1)$, $\omega_2(t_2)$, ..., $\omega_N(t_N)$ are canonical variables, in the vacuum state $|\Omega\rangle$
can be found in the interaction picture as
\beqa
&& T\langle \Omega|\omega_1(t_1) \omega_2(t_2)\cdots\omega_N(t_N)|\Omega\rangle \nonumber \\
& & = \frac{1}{Z}T \langle 0|\omega^{(I)}_1(t_1)\omega^{(I)}_2(t_2)\cdots\omega^{(I)}(t_N) \nonumber \\
&& \  \ \ \ \times \exp\left( - i \int_{-\infty}^\infty 
H_I(t)\,dt\right)|0\rangle\eqend{,} \label{eq:perturbative-omega}
\eeqa
where $H_I(t)$ is the interaction Hamiltonian and $\omega^{(I)}_i(t_i)$, $i=1,2,\ldots,N$, are the canonical variables $\omega_i(t_i)$ in the interaction picture.  Thus, the operators $\omega_i^{(I)}(t_i)$ satisfy the free equations.  Here, the state $|0\rangle$ is the tree-level vacuum state and
\beq
Z = T\langle 0| \exp\left( - i \int_{-\infty}^\infty H_I(t)dt\right)|0\rangle\eqend{.}
\eeq

The interacting vacuum state $|\Omega\rangle$ in de~Sitter spacetime cannot be defined in the same way as in Minkowski spacetime since the integral in Eq.~\eqref{eq:perturbative-omega} would be divergent due to the exponential growth of the space to the future and past.  Instead, we propose to define it so that the time-ordered $N$-point functions are the analytic continuation of those in the Euclidean theory obtained by the coordinate transformation \eqref{eq:Euclidean-time}.  Thus, for the Euclidean time the path-ordered product in the order of decreasing imaginary
part of $t$ to the left is defined by
\beqa
&& \mathcal{P}\langle \Omega|\omega_1(t_1) \omega_2(t_2)\cdots\omega_N(t_N)|\Omega\rangle \nonumber \\
& & = \frac{1}{Z_{\textrm{PE}}}
\mathcal{P} \langle 0|\omega^{(I)}_1(t_1)\omega^{(I)}_2(t_2)\cdots\omega^{(I)}(t_N) \nonumber \\
&& \  \ \ \ \ \ \ \ \ \ \ \times \exp\left( -  \int_{0}^\pi
H_I(t)\,d\tau\right)|0\rangle\eqend{,} \label{eq:Euclidean-Hamiltonian}
\eeqa
where
\beq
Z_{\textrm{PE}} = \mathcal{P} \langle 0|\exp\left( -  \int_{0}^{\pi} H_I(t)d\tau\right)|0\rangle\eqend{.}
\eeq
The path-ordering of operators such that the imaginary part of $t$ decreases to the left corresponds to the ordering such that the variable $\tau$ increases to the left.  The analytic continuation of the $N$-point functions in Eq.~\eqref{eq:Euclidean-Hamiltonian} to the real-time variables is performed by deforming the time-path as in the usual Schwinger-Keldysh perturbation theory~\cite{Schwinger:1960qe,Keldysh:1964ud} (see, e.g.~\cite{Landsman:1986uw}). This analytic continuation appears to be a concrete realization of the vacuum state proposed by Jacobson~\cite{Jacobson:1994fp} in the context of general spacetimes with bifurcate Killing horizons, which include de~Sitter spacetime.

The expectation value of the path-ordered product in Eq.~\eqref{eq:Euclidean-Hamiltonian} can be expressed as an integral of a product of two-point functions in the interaction picture by using Wick's theorem as in Lagrangian perturbation theory.  Thus, since $\langle 0|\omega Q_A^{(X)}|0\rangle = 0$, $X=H,c,\cbar$, where $\omega$ is any canonical variable, which is not any of the ``gauge-dependent variables'' $B_{(0)}^A$, $\theta^A$ or $\bar{\theta}^A$, at tree level,  we have $\langle \Omega|\Lambda Q_A^{(X)}|\Omega\rangle=0$ for any string $\Lambda$ of canonical variables not including $B_{(0)}^A$ $\theta^A$ or $\bar{\theta}^A$. That is, $Q_A^{(X)}|\Omega\rangle = 0$.

\section{Summary and Discussion}                                                                                                                    %
\label{sec:summary}                                                                                                                                 %

In this paper we found that there are conserved charges associated with the Killing vectors in perturbative gravity in the Landau gauge in spacetimes with compact Cauchy surfaces.  Our particular interest was perturbative quantum gravity in global de Sitter spacetime, where the de-Sitter-invariant FP-ghost propagator is IR divergent.  We propose that the physical states, in particular the vacuum state, should be annihilated by these charges.  Then we showed, assuming a certain definition of the vacuum state, that the use of the regularized de-Sitter-invariant FP-ghost propagator corresponds to requiring that the vacuum state be annihilated by the conserved charges mentioned above. (We note that this correspondence follows as long as the vacuum state is defined in such a way that the $N$-point function in Hamiltonian perturbation theory is obtained as an integral of a sum of products of the free-field two-point functions.)  Since the graviton propagator in global de~Sitter spacetime is IR finite~\cite{Faizal:2011iv} and that our FP-ghost propagator is effectively IR finite, we have a perturbation theory for quantum gravity in global de~Sitter spacetime which is not plagued by IR divergences coming from those in the propagators.

We also found that the BRST transforms of the charges $Q^{(\cbar)}_A$ are the conserved charges associated with the background spacetime symmetries.  (Although the gauge-fixing and FP-ghost terms break the general covariance, the gauge-fixed perturbative gravity action is still invariant under the background symmetries.)  Hence, a state annihilated by $Q^{(\cbar)}_A$ must be de~Sitter invariant in the case of de~Sitter spacetime.  The vacuum state is naturally de~Sitter invariant, but, since we propose that all physical states be annihilated by $Q^{(\cbar)}_A$ (and the other conserved charges $Q^{(c)}_A$ and $Q^{(H)}_A$), we must require also that they be de~Sitter invariant.  This condition is reminiscent of the quantum linearization stability conditions arising in linearized gravity quantized with the Dirac quantization method~\cite{Moncrief:1979bg,Moncrief:1978te,Higuchi:1991tk}.  Non-vacuum de~Sitter invariant states have been constructed using ``group-averaging'' to implement these conditions in Ref.~\cite{Higuchi:1991tm}.  We expect that the same method can be applied for the physical-state conditions in this paper.

In our analysis, the gauge parameter $\beta$ takes any real value not in a certain set of discrete values---see statement below Eq.~(\ref{eq:beta}). Due to the form of the gauge-fixing condition~(\ref{eq:gf-Lag}), $\beta$ only affects the scalar sectors of the graviton and the FP ghost fields. As mentioned above, for $\beta > 0$ the masses of these scalar fields are positive, making those sectors free of IR problems. For $\beta < 0$, but outside that discrete set, the scalar sectors are free of IR divergences, even though the propagators in the de~Sitter-invariant vacuum grow at large point separations. In principle, this growth can spoil the convergence in the IR of the Feynman diagrams making the perturbative series of gauge-dependent correlators.\footnote{On general grounds, we do not expect this to be a problem for gauge-invariant correlators as they should not depend on gauge parameters.} This problem, however, is avoided in the perturbation theory laid out in Sec.~\ref{sec:Hamiltonian}, as the diagram vertices are integrated over the Euclidean section of de~Sitter spacetime, the $n$-sphere. The deformation for the Euclidean contour into the Schwinger-Keldysh contour will not change this, as long as we keep the initial time finite, since the Schwinger-Keldysh is causal and the spatial section of de~Sitter spacetime is the $(n - 1)$-sphere. As for the values of $\beta$ in the set $-s(s + n - 1)/(n - 1)$, with $s = 0,1,2,\dots$, the scalar sectors of the graviton and FP ghosts have IR divergences similar to the massless case discussed in Sec.~\ref{sec:Killing-vector-modes}, and corresponds to the tachyonic fields of Ref.~\cite{Bros:2010wa}. Although it could be interesting to try to accommodate them in the framework we have discussed, these discrete values are not associated with any particularly relevant gauge condition in four dimensions.

It is interesting to compare our proposal to deal with the ghost zero modes problem with the approach of Refs.~\cite{Polchinski:1988ua, Taylor:1989ua, Vassilevich:1992rk}. There, the one-loop effective action was computed by restricting the FP determinant to the modes with strictly positive eigenvalues corresponding to the Euclidean vacuum. In our work, the condition that the physical state is annihilated by $Q^{(c)}_A$ and $Q^{(\cbar)}_A$ implies, at tree level, that the operators $c^A_{(0)}$ and $\cbar^A_{(0)}$, defined in Eq.~(\ref{eq:zero_mode_projection}), are effectively time-independent when acting on the state. This condition, however, makes the zero-modes state non-normalizable, and thus the next step is to change the inner product for the space of states so it does not involve the zero modes---see Sec.~\ref{sec:Killing-vector-modes}. In the path-integral formalism, the change of the inner product consists in removing the integration over the zero modes from the functional integral. This is precisely the method employed in Refs.~\cite{Polchinski:1988ua, Taylor:1989ua, Vassilevich:1992rk}. The advantage of our proposal is that it allows us to consistently remove the zero modes beyond the one-loop order, and to show that their removal is compatible with the BRST symmetry of the gauge-fixed action, i.e. it does not spoil unitarity. 

Our definition of the vacuum state involves imaginary time but it uses the Hamiltonian rather than the Lagrangian. The Euclidean action obtained as the integral of the Einstein-Hilbert Lagrangian over an Euclidean section is not bounded from below (the conformal-mode problem~\cite{Gibbons:1978ac,Schleich:1987fm}). This problem is usually dealt with via a ``conformal rotation'', which consists of changing the sign of the kinetic term of the conformal mode. It would be interesting to investigate how this problem manifests itself in our definition of the vacuum state in Hamiltonian perturbation theory.  Finally, it would be interesting to investigate whether our proposal for the physical states in gauge theory and perturbative gravity gives any insight into the discrepancy between the unitary and covariant formulations of these theories in de~Sitter spacetime, or in curved spacetime in general~\cite{Vassilevich:1991rt,Vassilevich:1992ad,Fukutaka:1992si,Vassilevich:1994cz,Donnelly:2013tia}.

\acknowledgments

This work was supported by the grant  no.~RPG-2018-400, ``Euclidean and in-in formalisms in static spacetimes with Killing horizons'', from the Leverhulme Trust.  The work of J.G.\ was supported by a Studentship from the Engineering and Physical Sciences Research Council (EPSRC).

\appendix

\section{Derivation of Eq.~(\ref{newlagrangian})}                                                                                                   %
\label{appA}                                                                                                                                        %

In this Appendix we derive the form of the Lagrangian density \eqref{newlagrangian} convenient for finding the conserved charges involving the antighost field. It is convenient to consider the Lagrangian density obtained from $\mathcal{L}_{\textrm{FP}}$ by interchanging the roles of the ghost and antighost fields:
\begin{eqnarray}
\mathcal{L}_{\overline{\textrm{FP}}}
& \equiv & -i(\nabla_\mu\cbar_\nu + \nabla_\nu \cbar_\mu - 2kg_{\mu\nu}\nabla_\lambda \cbar^\lambda)
\nabla^\mu c^\nu
\nonumber \\
&& - i \pounds_{\cbar}h_{\mu\nu} \nabla^\mu c^\nu + ik g^{\beta\gamma}\pounds_{\cbar} h_{\beta\gamma}
\nabla_\alpha c^\alpha\eqend{.}
\end{eqnarray}
We find
\begin{eqnarray}
\mathcal{L}_{\textrm{FP}} - \mathcal{L}_{\overline{\textrm{FP}}} 
& = & -i\left[ (\nabla^\mu \cbar^\nu) c^\alpha  - \cbar^\alpha (\nabla^\mu c^\nu) \right] \nabla_\alpha H_{\mu\nu}
\nonumber \\
&&  - i\left[ 
\nabla^\mu \cbar^\nu\nabla_\nu c^\alpha - \nabla_\nu \cbar^\alpha \nabla^\mu c^\nu\right] H_{\alpha\mu} \nonumber \\
&& + 
2ik \left[ \nabla_\alpha \cbar^\alpha \nabla_\mu c_\nu  - \nabla_\mu \cbar_\nu \nabla_\alpha c^\alpha\right]
H^{\mu\nu}\eqend{,} \nonumber \\
\end{eqnarray}
where we recall that $H_{\mu\nu} \equiv h_{\mu\nu} - k g_{\mu\nu} h$. Next, we ``integrate by parts'' the first term to remove the derivative $\nabla_\alpha$ on $H_{\mu\nu}$, and then commute the derivatives $\nabla^\mu$ and $\nabla_\alpha$.  The terms containing the Riemann tensors arising  from this procedure cancel out.  Thus we obtain
\begin{equation}
\mathcal{L}_{\textrm{FP}} = \mathcal{L}_{\overline{\textrm{FP}}} 
+ K_{\mu\nu}H^{\mu\nu}\eqend{,} \label{L-FP-bar}
\end{equation}
up to a total divergence, where
\begin{eqnarray}
K_{\mu\nu} & \equiv & 
 i \nabla_\mu \left[ (\nabla_\alpha \cbar_\nu )c^\alpha - \cbar^\alpha \nabla_\alpha c_\nu\right] \nonumber \\
&& 
+ i(1-2k)\left[\nabla_\mu \cbar_\nu \nabla_\alpha c^\alpha - \nabla_\alpha\cbar^\alpha \nabla_\mu c_\nu\right]\eqend{.}
 \end{eqnarray}
Then, we find Eq.~\eqref{newlagrangian} by 
adding $\mathcal{L}_{\textrm{gf}} = - \nabla^\mu B^\nu H_{\mu\nu}$ to the right-hand side of 
Eq.~\eqref{L-FP-bar}.

\section{BRST Transform of the Charge \texorpdfstring{$Q_A^{(\bar{c})}$}{Eq.~(\ref{charge-cbar})}}                                                  %
\label{AppB}                                                                                                                                        %

In this Appendix we show that the charge $\delta_{\textrm{B}} Q_A^{(\bar{c})}$ for each $A$ is proportional to the Noether charge for the spacetime symmetry generated by the Killing vector $\xi_{A}^\mu$.  We write $\xi_A^\mu$ simply as $\xi^\mu$, dropping the subscript $A$, in the rest of this Appendix. 

First, we find the BRST transform of the conserved current $\xi_{\nu} \mathcal{T}^{\mu\nu}$ which corresponds to the conserved charge defined by Eq.~\eqref{charge-cbar}. Let us write
\beq
\delta_{\textrm{B}} (\xi^\nu\mathcal{T}_{\mu\nu}) = J^{(\rB,B)}_\mu + J^{(\rB,\bar{c}c)}_\mu\eqend{,}
\label{BRST-tr-of-T}
\eeq
where $J^{(\rB,B)}_\mu$ comes from the BRST transformation of $\bar{c}^\alpha$ whereas
 $J^{(\rB,\bar{c}c)}_\mu$
comes from that of $h_{\mu\nu}$. (Recall that $\delta_{\textrm{B}} B^\mu = 0$.)  The current $J_\mu^{(\rB,B)}$ is obtained by replacing $\bar{c}^\alpha$ by $iB^\alpha$ in $\xi_{\nu} \mathcal{T}^{\mu\nu}$. 
It is convenient to write
\beqa
\mathcal{T}_{\mu\nu}
& = &  T_{\mu\nu} + \pounds_{\bar{c}}H_{\mu\nu}  
+ k T_{\mu\nu} h  
\nonumber \\
&&  -g_{\mu\nu} \nabla^\alpha \bar{c}^\beta H_{\alpha\beta} + (1-2k) \nabla_\alpha \bar{c}^\alpha H_{\mu\nu}
\eqend{,} 
\label{convenient-T}
\eeqa
where we have defined
\beq
T_{\mu\nu} \equiv \nabla_\mu \bar{c}_\nu + \nabla_\nu \bar{c}_\mu 
- 2kg_{\mu\nu}\nabla_\alpha\bar{c}^\alpha\eqend{.} \label{eq:def-of-Tmunu}
\eeq
Thus,
\beqa
J^{(\rB,B)}_\mu
& = & i \xi^\nu \left[ (\nabla_\mu B_\nu
+ \nabla_\nu B_\mu - 2k g_{\mu\nu}\nabla_\alpha B^\alpha)(1+kh) \right. \nonumber \\
&& \left.
+ \pounds_B H_{\mu\nu}  - g_{\mu\nu} \nabla^\alpha B^\beta H_{\alpha\beta}\right. \nonumber \\
&& \left. + (1-2k)H_{\mu\nu} \nabla_\alpha B^\alpha\right]\eqend{.} 
\label{eq:JBB}
\eeqa
To find the part of the current coming from the transformation of $h_{\mu\nu}$ we use
\begin{subequations}
\beqa
\delta_{\textrm{B}} H_{\mu\nu} & = & \mathcal{S}_{\mu\nu}\eqend{,}\\
\delta_{\textrm{B}} h & = & 
2 \nabla_\alpha c^\alpha + g^{\alpha\beta}\pounds_c h_{\alpha\beta}\eqend{.}
\eeqa
\end{subequations}
We note that there will be an extra minus sign in the transformation of $\mathcal{T}_{\mu\nu}$ in Eq.~\eqref{convenient-T} because $H_{\mu\nu}$ and $h$ are to the right of a fermionic variable $\bar{c}^\alpha$. Thus, we find this part of the current as
\beqa
J^{(\rB,\bar{c}c)}_\mu
& = & - \xi^\alpha\left[ \bar{c}^\beta \nabla_\beta \mathcal{S}_{\mu\alpha}
+ \nabla_\mu \bar{c}^\beta \mathcal{S}_{\beta\alpha} + \nabla_\alpha\bar{c}^\beta \mathcal{S}_{\beta\mu}
\right. \nonumber \\
&& - g_{\mu\alpha}\nabla^\beta \bar{c}^\gamma \mathcal{S}_{\beta\gamma}
 + (1-2k)\nabla_\beta \bar{c}^\beta \mathcal{S}_{\mu\alpha} \nonumber \\
&& \left. + k T_{\mu\alpha}\left( 2\nabla_\lambda c^\lambda
+ g^{\beta\gamma}\pounds_c h_{\beta\gamma}\right) \right]\eqend{.}  \label{Jcc}
\eeqa

Next, we construct the conserved current for the spacetime symmetry generated by $\xi^\mu$.  Consider the diffeomorphism transformation given by
\begin{subequations}  \label{eq:spacetime-transformation}
\begin{eqnarray}
\delta_{\textrm{st}} h_{\mu\nu} & = & \nabla_\mu (\alpha\xi_\nu) + \nabla_\nu (\alpha\xi_\mu) 
+ \pounds_{\alpha \xi}h_{\mu\nu} \nonumber \\
& = & \nabla_\mu (\alpha\xi_\nu) + \nabla_\nu (\alpha\xi_\mu) + 
\alpha \xi^\lambda \nabla_\lambda h_{\mu\nu} \nonumber \\
&& + \nabla_\mu (\alpha \xi^\lambda)h_{\lambda \nu}
+ \nabla_\nu (\alpha \xi^\lambda)h_{\mu\lambda},\label{h-transform}\\
\delta_{\textrm{st}} \psi^\mu & = & \pounds_{\alpha\xi} \psi^\mu \nonumber \\
& = & \alpha \left[ \xi^\lambda\nabla_\lambda \psi^\mu - (\nabla_\lambda \xi^\mu)\psi^\lambda\right]
- (\nabla_\lambda\alpha) \xi^\mu \psi^\lambda\eqend{,} \nonumber \\
\end{eqnarray}
\end{subequations}
where $\psi^\mu = B^\mu$, $c^\mu$ or $\bar{c}^\mu$ and where $\alpha$ is a compactly-supported function on the background spacetime. Let us represent the transformation~\eqref{eq:spacetime-transformation} as
\beq
\delta_{\textrm{st}} \Phi_I = \alpha X_I + Y_I^\lambda \nabla_\lambda \alpha\eqend{,} \label{Phi-I-transf}
\eeq
where $\Phi_I$ represents $h_{\mu\nu}$, $B^\mu$, $c^\mu$ or $\bar{c}^\mu$ depending on the index $I$. Let $\mathcal{L} \equiv \mathcal{L}_{\textrm{FP}} + \mathcal{L}_{\textrm{gf}}$.  Then
\beqa
\delta_{\textrm{st}} \mathcal{L}
& = & \left[  \frac{\partial\mathcal{L}}{\partial\Phi_I}
-  \nabla_\mu \left(\frac{\partial\mathcal{L}}{\partial (\nabla_\mu \Phi_I)}\right)\right]
(\alpha X_I + Y_I^\lambda \nabla_\lambda \alpha)\eqend{,} \nonumber \\
\label{delta-L-1}
\eeqa
where we have dropped a total divergence. In this equation the index $I$ is summed over. Since the transformation \eqref{Phi-I-transf} with constant $\alpha$ would give the spacetime symmetry transformation generated by the Killing vector $\xi^\mu$, we have
\beqa
\delta_{\textrm{st}}\mathcal{L}|_{\alpha=\textrm{const.}}
& = & \nabla_\mu (\xi^\mu \mathcal{L}) \nonumber \\
& = &  \frac{\partial\mathcal{L}}{\partial\Phi_I}X_I
+ \left(\frac{\partial\mathcal{L}}{\partial (\nabla_\mu \Phi_I)}\right)\nabla_\mu X_I\eqend{.}
\eeqa
By using this equation in Eq.~\eqref{delta-L-1} we obtain
\beqa
\delta_{\textrm{st}} \mathcal{L}
& = & 
\left\{  \left(\frac{\partial\mathcal{L}}{\partial (\nabla_\mu \Phi_I)}\right)X_I  - \xi^\mu \mathcal{L} \right. 
\nonumber \\
&&\left. +  \left[  \frac{\partial\mathcal{L}}{\partial\Phi_I}
-  \nabla_\lambda \left(\frac{\partial\mathcal{L}}{\partial (\nabla_\lambda \Phi_I)}\right)\right]Y_I^\mu \right\}
\nabla_\mu \alpha\eqend{,} \label{eq:what-is-this}
\eeqa
with a total divergence dropped. 

Now, let $\mathcal{L}_{\textrm{GR}} = \sqrt{-g}\,\mathcal{L}_{\textrm{EH}}$ be the standard Einstein-Hilbert Lagrangian density for gravity. Then, since $\mathcal{L}_{\textrm{EH}} + \mathcal{L}$ is the total Lagrangian density, $\delta_{\textrm{st}}\mathcal{L}_{\textrm{EH}} + \delta_{\textrm{st}}\mathcal{L}$ must be a total divergence if the field equations are satisfied.  But $\delta_{\textrm{st}}\mathcal{L}_{\textrm{EH}}$ under the transformation \eqref{h-transform} is a total divergence (even if the field equations are not satisfied) since the corresponding Einstein-Hilbert action is diffeomorphism invariant. Hence, $\delta_{\textrm{st}}\mathcal{L}$ must be a total divergence if the field equations are satisfied. This implies that the expression inside the curly brackets in Eq.~\eqref{eq:what-is-this} must be divergence-free if the field equations are satisfied. Hence we identify the spacetime-symmetry current as
\beqa
J_{\textrm{st}}^\mu & = & 
 \left(\frac{\partial\mathcal{L}}{\partial (\nabla_\mu \Phi_I)}\right)X_I  - \xi^\mu \mathcal{L}  \nonumber \\
&&  +  \left[  \frac{\partial\mathcal{L}}{\partial\Phi_I}
-  \nabla_\lambda \left(\frac{\partial\mathcal{L}}{\partial (\nabla_\lambda \Phi_I)}\right)\right]Y_I^\mu\eqend{.}
\label{current-general}
\eeqa
Note that the term proportional to $Y_I^\mu$ is absent for $\Phi_I = B^\mu$, $c^\mu$ or $\bar{c}^\mu$ because it is proportional to the field equation for $\Phi_I$ in these cases.  This is not the case for $\Phi_I = h_{\mu\nu}$, however, because its field equation comes from $\mathcal{L}_{\textrm{EH}} + \mathcal{L}$, not just from $\mathcal{L}$.

Let us find the part of the current $J_{\textrm{st}}^\mu$ given by Eq.~\eqref{current-general} coming from $\mathcal{L}_{\textrm{gf}}$. Since $\mathcal{L}_{\textrm{gf}}$ depends on $\nabla_\mu B_\nu$ and $h_{\mu\nu}$ but not on $B^\mu$ or $\nabla_\alpha h_{\mu\nu}$, this part of the current is
\begin{splitequation}\label{B-current}
J^{(B)\mu}_{\textrm{st}}
& = \frac{\partial \mathcal{L}_{\textrm{gf}}}{\partial h_{\alpha\beta}} \left[  \delta^\mu_\alpha \xi_\beta + \delta^\mu_\beta \xi_\alpha + 
\delta^\mu_\alpha \xi^\lambda h_{\lambda\beta} + \delta^\mu_\beta \xi^\lambda h_{\lambda \alpha} \right]\\
& \phantom{=}\; + \frac{\partial \mathcal{L}_{\textrm{gf}}}{\partial (\nabla_\mu B_\nu)}\left[ \xi^\lambda \nabla_\lambda B_\nu - (\nabla_\lambda \xi_\nu)B^\lambda\right]\\
& \phantom{=}\; + \xi^\mu \nabla^\alpha B^\beta H_{\alpha\beta}\\
& = - (\xi_\alpha + H_{\alpha\lambda}\xi^\lambda + k h\xi_\alpha)(\nabla^\mu B^\alpha + \nabla^\alpha B^\mu \\
& \phantom{=}\; - 2k g^{\mu\alpha}\nabla_\beta B^\beta) - H^{\mu\nu}\left[ \xi^\lambda \nabla_\lambda B_\nu - (\nabla_\lambda \xi_\nu)B^\lambda\right]\\
& \phantom{=}\; + \xi^\mu \nabla^\alpha B^\beta H_{\alpha\beta}\eqend{.}
\end{splitequation}
Then we find
\beq
J^{(B)\mu}_{\textrm{st}} =  i J^{(\rB,B)\mu} +  \nabla_\nu F^{(1)\mu\nu} \eqend{,}
\label{B-equality}
\eeq
where $J^{(\rB,B)\mu}$ is given by Eq.~\eqref{eq:JBB} and
\beq
F^{(1)\alpha\mu} \equiv B^\mu \xi_\beta H^{\alpha\beta} - B^\alpha \xi_\beta H^{\mu\beta}\eqend{,}
\eeq
which is an antisymmetric tensor. 

Next, the part of the current $J_{\textrm{st}}^\mu$ coming from the variation of $\mathcal{L}_{\textrm{FP}}$ with respect to
$\bar{c}_\nu$ plus the term $-\xi^\mu \mathcal{L}_{\textrm{FP}}$ reads
\begin{splitequation}
J_{\textrm{st}}^{(\bar{c}c, \bar{c})\mu}
& = - i (\xi^\alpha \nabla_\alpha\bar{c}^\beta {\mathcal{S}^\mu}_\beta - \nabla_\alpha \xi^\beta \bar{c}^\alpha {\mathcal{S}^\mu}_\beta\\
& \phantom{=}\; - \xi^\mu \nabla^\beta \bar{c}^\gamma \mathcal{S}_{\beta\gamma})\eqend{,} \\
\end{splitequation}
where $\mathcal{S}_{\mu\nu}$ is defined by Eq.~\eqref{mathcalC}.   Define
\beq
F^{(2)\mu\nu} \equiv i \xi^\alpha \left( \cbar^\mu {\mathcal{S}^\nu}_\alpha - \cbar^\nu {\mathcal{S}^\mu}_\alpha
\right)
\eqend{,}
\eeq
which is an antisymmetric tensor.  Then, we find, using the field equation $\nabla_\alpha \mathcal{S}^{\alpha\beta}=0$ and the antisymmetry of the tensor $\nabla_\alpha \xi_\beta$,
\beqa
J_{\textrm{st}}^{(\bar{c}c, \bar{c})\mu}
& = & -i \left[\xi^\alpha\nabla_\alpha \bar{c}^\beta {\mathcal{S}^\mu}_\beta 
- \xi^\mu \nabla^\beta \bar{c}^\gamma \mathcal{S}_{\beta\gamma}\right. \nonumber \\
&& \left. + \xi^\beta \nabla_\alpha (\bar{c}^\alpha {\mathcal{S}^\mu}_\beta) 
- \xi^\beta \nabla_\alpha \bar{c}^\mu {\mathcal{S}^\alpha}_\beta\right]\eqend{,} \nonumber \\
&& + \nabla_\nu F^{(2)\mu\nu}\eqend{.}
\label{cbar-current}
\eeqa

The part of $J_{\textrm{st}}^\mu$ coming from the variation of $\mathcal{L}_{\textrm{FP}}$ with respect to $c^\alpha$ reads
\beqa
\tilde{J}^{(\bar{c}c,c)}_{st\,\mu} & = & 
-i T_{\mu\nu} (\xi^\alpha \nabla_\alpha c^\nu - \nabla_\alpha \xi^\nu c^\alpha) \nonumber  \\
&& 
- i T_{\mu\beta}h^{\beta\nu} (\xi^\alpha \nabla_\alpha c_\nu - \nabla_\alpha \xi_\nu c^\alpha)\eqend{,}
\label{c-current}
\eeqa
where $T_{\mu\nu}$ is given by Eq.~\eqref{eq:def-of-Tmunu}. The contribution from varying $h_{\mu\nu}$ in $\frac{1}{2} T^{\mu\nu} c^\alpha \nabla_\alpha h_{\mu\nu}$ is
\begin{eqnarray}
\tilde{J}^{(\bar{c}c,h1)\mu}_{\textrm{st}}
& = & i \nabla_\alpha (T^{\mu\nu}c^\alpha)(\xi_\nu + \xi^\beta h_{\beta\nu}) \nonumber \\
&& -  \frac{i}{2}T^{\beta\gamma}c^\mu \xi^\alpha \nabla_\alpha h_{\beta\gamma} 
 - i T^{\alpha \beta}c^\mu (\nabla_\alpha \xi^\lambda) h_{\lambda \beta}\eqend{.} \nonumber \\
\label{h1-current}
\end{eqnarray}
Then we find
\beqa
&& J^{(\bar{c}c,c)\mu}_{\textrm{st}} + J^{(\bar{c}c,h1)\mu}_{\textrm{st}}\nonumber \\
 && =  -iT^{\mu\nu}\xi^\alpha \nabla_\alpha c_\nu
- i T^{\mu\beta}h_{\beta\nu}\xi^\alpha \nabla_\alpha c^\nu \nonumber \\
&& \ \ - \frac{i}{2}T^{\beta\gamma}c^\mu \xi^\alpha \nabla_\alpha h_{\beta\gamma} 
- i \xi^\nu T^{\mu\beta}c^\alpha \nabla_\alpha h_{\beta\nu} \nonumber \\
&& \ \ + i \xi_\nu \nabla_\alpha\left[ (T^{\alpha\nu} + T^{\alpha\beta}{h_\beta}^\nu)c^\mu\right]
+  \nabla_\nu F^{(3)\mu\nu}\eqend{,}
\label{almost}
\eeqa
where the antisymmetric tensor $F^{(3)\mu\nu}$ is given by
\beq
F^{(3)\mu\nu} \equiv i\xi_\lambda \left[ (T^{\mu\lambda} + T^{\mu\beta}{h_\beta}^\lambda)c^\nu
- (T^{\nu\lambda} + T^{\nu\beta}{h_\beta}^\lambda)c^\mu\right]\eqend{.}
\eeq
Next, we note that the field equation obtained by varying the action with the Lagrangian density $\mathcal{L} = \mathcal{L}_{\textrm{FP}} + \mathcal{L}_{\textrm{gf}}$ with respect to $c^\alpha$ can be written as
\beq
\nabla_\beta \left[ T^{\beta\alpha} + T^{\beta\nu}{h^\alpha}_\nu \right]
 =  \frac{1}{2}T^{\nu\beta} \nabla^\alpha h_{\beta\nu}\eqend{.}
\eeq
(This equation can also be derived from $\nabla_\nu \mathcal{T}^{\mu\nu} = 0$ and $\nabla_\nu H^{\mu\nu}=0$.)  By using this equation in Eq.~\eqref{almost} we find
\begin{splitequation}
& J^{(\bar{c}c,c)\mu}_{\textrm{st}} + J^{(\bar{c}c,h1)\mu}_{\textrm{st}}\\
&\;\; =   -iT^{\mu\nu}\xi^\alpha \nabla_\alpha c_\nu - i T^{\mu\beta}h_{\beta\nu}\xi^\alpha \nabla_\alpha c^\nu - i \xi^\nu T^{\mu\beta}c^\alpha \nabla_\alpha h_{\beta\nu} \\
&\;\;\phantom{=}\; + i\xi_\nu (T^{\alpha\nu} + T^{\alpha\beta}{h_\beta}^\nu)\nabla_\alpha c^\mu + \nabla_\nu F^{(3)\mu\nu}\eqend{.}
\end{splitequation}
Finally, the part coming from the variation of $h_{\mu\nu}$ in the term $-i T^{\mu\nu}\nabla_\nu c^\alpha h_{\alpha\nu}$ in
$\mathcal{L}_{\textrm{FP}}$ is
\begin{eqnarray}
J^{(\bar{c}c,h2)}_{\textrm{st}\,\mu}
& = & -i \xi^\alpha ({T^\beta}_\alpha \nabla_\beta c_\mu + T_{\beta\mu} \nabla^\beta c_\alpha \nonumber \\
&& \ \ \ + T^{\beta\gamma}\nabla_\beta c_\mu h_{\alpha\gamma} 
+ T_{\gamma\mu}\nabla^\gamma c^\beta h_{\alpha\beta})\eqend{.}
\label{h2-current}
\end{eqnarray}
Then
\beqa
J^{(\bar{c}c,ch)\mu}_{\textrm{st}} & \equiv & J^{(\bar{c}c,c)\mu}_{\textrm{st}} + J^{(\bar{c}c,h1)\mu}_{\textrm{st}}
+ J^{(\bar{c}c,h2)\mu}_{\textrm{st}} \nonumber \\
& = & -i \xi^\alpha T^{\mu\beta}\left(
\nabla_\alpha c_\beta + \nabla_\beta c_\alpha + \pounds_c h_{\alpha\beta}\right) \nonumber \\
&& + \nabla_\nu(F^{(2)\mu\nu}  + F^{(3)\mu\nu})\eqend{.} \label{ch-current}
\eeqa 

The Noether current for the spacetime symmetries generated by the Killing vector $\xi^\mu$ is
\beq
J_{\textrm{st}}^\mu = J_{\textrm{st}}^{(B)\mu} + J_{\textrm{st}}^{(\bar{c}c,\bar{c})\mu} +
J_{\textrm{st}}^{(\bar{c}c,ch)\mu}\eqend{,}
\label{total-spacetime-current}
\eeq
where the currents on the right-hand side are given by Eqs.~\eqref{B-current}, \eqref{cbar-current} and \eqref{ch-current}.  By a straightforward calculation we can show that
\beq
J_{\textrm{st}}^{(\bar{c}c,\bar{c})\mu} + J_{\textrm{st}}^{(\bar{c}c,ch)\mu}
= iJ^{(\rB,\bar{c}c)\mu} + \nabla_\nu(F^{(2)\mu\nu} + F^{(3)\mu\nu})\eqend{.}
\eeq
This equation and Eq.~\eqref{B-equality}, together with Eq.~\eqref{BRST-tr-of-T} and \eqref{total-spacetime-current}, imply that 
\beq
J_{\textrm{st}}^\mu = i \delta_{\textrm{B}}(\xi_\nu \mathcal{T}^{\mu\nu})
+ \nabla_\nu (F^{(1)\mu\nu} + F^{(2)\mu\nu} + F^{(3)\mu\nu})\eqend{.}
\eeq
Thus, by defining the Noether charges for the spacetime symmetries generated by the Killing vector $\xi_A^\mu$ by
\beq
Q^{(\textrm{st})}_A \equiv \int_{\Sigma}d\Sigma\, n_\mu J_{\textrm{st}}^\mu\eqend{,}
\eeq
with $\xi^\mu =\xi_A^\mu$, we indeed have $Q^{(\textrm{st})}_A = i \delta_{\textrm{B}}Q_A^{(\cbar)}$, by the generalized Stokes theorem.

\section{Derivation of Eq.~(\ref{eq:alternativeL-1})}                                                                                               %
\label{App:hamiltonian-approach}                                                                                                                    %

What we need to show is that the nonlinear terms in $\mathcal{L}_{\textrm{FP}}$ given by Eq.~\eqref{FP-Lag} equals the nonlinear terms in Eq.~\eqref{eq:alternativeL-1} with $\check{B}^\mu = \check{B}^\mu|_{B^\alpha=0} =  - i (\nabla_\alpha\bar{c}^\mu)c^\alpha$, where $\check{B}^\mu$ is defined  by Eq.~\eqref{eq:checkB}. That is, we need to show that the part involving $h_{\mu\nu}$ in $\mathcal{L}_{\textrm{FP}}$ plus $\nabla^\mu\check{B}^\nu|_{B^\alpha=0}H_{\mu\nu}$ equals the terms involving $h_{\mu\nu}$ in Eq.~\eqref{eq:alternativeL-1}. The former reads
\begin{splitequation}
\mathcal{L}_{\textrm{FP}}^{(h)} 
& = - i \nabla^\mu \bar{c}^\nu[c^\alpha \nabla_\alpha H_{\mu\nu} + \nabla_\mu c^\alpha H_{\alpha\nu} + \nabla_\nu c^\alpha H_{\alpha\mu} \\
&\phantom{=}\; + k \left(\nabla_\mu c_\nu + \nabla_\nu c_\mu - 2k g_{\mu\nu}\nabla_\alpha c^\alpha\right)h] \\
&\phantom{=}\; - i\nabla^\mu\left[ (\nabla_\alpha\bar{c}^\nu)c^\alpha\right] H_{\mu\nu}\eqend{.} \label{eq:FPh}
\end{splitequation}
The first term contains the time derivative of $H_{\mu\nu}$. It can be combined with the last term as
\beqa
&&  - i (\nabla^\mu \bar{c}^\nu)
 c^\alpha \nabla_\alpha H_{\mu\nu}   
- i\nabla^\mu\left[ (\nabla_\alpha\bar{c}^\nu)c^\alpha\right] H_{\mu\nu} \nonumber \\
&& = - i\nabla_\alpha\left[ (\nabla^\mu \bar{c}^\nu)c^\alpha H_{\mu\nu}\right] 
 + i\nabla_\alpha\left[ (\nabla^\mu\bar{c}^\nu)c^\alpha\right]H_{\mu\nu} \nonumber \\
&& \ \ \   -i
\big\{
\left[ (\nabla_\alpha \bar{c}^\nu)\nabla^\mu c^\alpha + (\nabla_\alpha \nabla^\mu \bar{c}^\nu)c^\alpha
- {{R^\nu}_{\beta\alpha}}^\mu \bar{c}^\beta c^\alpha\right]\big\}H_{\mu\nu}\eqend{.} \nonumber \\
\eeqa
By substituting this formula into Eq.~\eqref{eq:FPh} we find  after some simplification that 
$\mathcal{L}^{(h)}_{\textrm{FP}}$ is equal to the terms involving $h_{\mu\nu}$ in Eq.~\eqref{eq:alternativeL-1}
up to a total divergence.

\section{The Bosonic Condition on the Vacuum State at Tree Level}                                                                                   %
\label{non-contrib}                                                                                                                                 %

The solutions to the field equations for the gravitons at linearized level have been studied in de~Sitter spacetime in global coordinates in  Ref.~\cite{Faizal:2011iv}.  The graviton field with a small mass term is expressed as $h_{\mu\nu} = h_{\mu\nu}^{(T)} + h_{\mu\nu}^{(V)} + h_{\mu\nu}^{(S)}$. The tensor sector $h_{\mu\nu}^{(T)}$ contains the mode functions composed of the tensor, vector or scalar spherical harmonics with angular momentum $\ell \geq 2$.  This implies, by orthogonality of spherical harmonics with respect to the space integral, that $h_{\mu\nu}^{(T)}$ does not contribute to the conserved charge $Q^{(H)}_A$. The reason for this is that $Q^{(H)}_A$ is defined by a space integral of the product of $h_{\mu\nu}$ and the Killing vector $\xi_A^\mu$ composed of the scalar or vector spherical harmonic with $\ell=1$. (Although linearized gravity was studied only in $4$ dimensions in Ref.~\cite{Faizal:2011iv} these facts hold in $n$ dimensions as well.) 

The conserved bosonic charge $Q^{(H)}_A$ has no contribution from the scalar sector, either. To show this, we first express the scalar sector $h_{\mu\nu}^{(S)}$ as~\cite{Frob:2016hkx}
\beq
 h_{\mu\nu}^{(S)} = \nabla_\mu \nabla_\nu \Phi + g_{\mu\nu}\Psi\eqend{,} \label{eq:scalar-sector}
\eeq
where  
\begin{subequations}
\beqa
\Box \Phi  & = & (n-1)\beta\Phi - \left[n+(n-1)\beta - \frac{n-2}{2}\alpha \beta\right]\Psi\eqend{,} 
\nonumber \\ \label{eq:Phi-eq}\\
\Box \Psi & = & (n-1)\beta \Psi\eqend{,}  \label{eq:Psi-equation}
\eeqa
\end{subequations}
in the covariant gauge given by the gauge-fixing term~\eqref{eq:new-gf-term}.  
We will show that the scalar contribution to the charge $Q_A^{(H)}$,
\beq
Q_A^{(H,S)} \equiv \int_\Sigma d\Sigma n_\mu \xi_{A\nu}\left( h^{(S)\mu\nu} - k g^{\mu\nu} h^{(S)}\right) 
\eqend{,}  \label{eq:charge-S-integral}
\eeq
vanishes in the Landau-gauge limit $\alpha\to 0$.

By Eqs.~ \eqref{eq:scalar-sector} and \eqref{eq:Phi-eq} we find
\beqa
h^{(S)}_{\mu\nu} - kg_{\mu\nu}h^{(S)}
& = & \nabla_\mu\nabla_\nu \Phi - g_{\mu\nu} (\Box + n-1)\Phi \nonumber \\
&& + g_{\mu\nu}\frac{n-2}{2}\alpha \Psi \eqend{.}
\eeqa
Then, by using $\Box \xi^\mu_A = - (n-1)\xi^\mu_A$, which readily follows from the identities
$\nabla_\rho \nabla_\mu \xi^\nu_A = {R^\nu}_{\mu\rho\sigma}\xi_A^\sigma$ and
$R_{\nu\mu\rho\sigma} = g_{\nu\rho}g_{\mu\sigma} - g_{\nu\sigma}g_{\mu\rho}$, we find
\beqa
&& \xi_{A\nu}( h^{(S)\mu\nu} - kg^{\mu\nu}h^{(S)}) \nonumber \\
& & = \xi^{\nu}_A\nabla_\nu \nabla^\mu \Phi - \xi^\mu_A \nabla_\nu \nabla^\nu \Phi 
 + (\nabla_\nu \nabla^\nu \xi_A^\mu)\Phi + \frac{n-2}{2}\alpha \xi_A^\mu \Psi \nonumber \\
& & =  \nabla_\nu\left[ \xi_A^\nu \nabla^\mu \Phi - \xi_A^\mu \nabla^\nu \Phi + (\nabla^\nu \xi^\mu_A)\Phi\right]
+ \frac{n-2}{2}\alpha \xi_A^\mu \Psi\eqend{.}
\eeqa 
Substituting this formula into Eq.~\eqref{eq:charge-S-integral} yields 
\beq
Q^{(H,S)}_A = \frac{n-2}{2}\alpha \int_{\Sigma}d\Sigma\,n_\mu \xi_A^\mu \Psi\eqend{,}
\eeq
after using the generalized Stokes theorem. Thus, $Q^{(H,S)}_A \to 0$ in the Landau-gauge limit $\alpha\to 0$.  [The scalar field $\Psi$, which satisfies the massive Klein-Gordon equation (\ref{eq:Psi-equation}), has a finite limit as $\alpha\to 0$.]  Therefore, only the vector sector contributes to the conserved charge $Q^{(H)}_A$ at tree level.

The vector sector of the graviton field can be expressed as
\beq
h_{\mu\nu}^{(V)} = \nabla_\mu V_\nu + \nabla_\nu V_\mu\eqend{,}
\eeq
where $\nabla^\rho V_\rho = 0$. Since the linearized Einstein-Hilbert action is invariant under linearized gauge transformations, $h_{\mu\nu}\to h_{\mu\nu}+\nabla_\mu \Lambda_\nu + \nabla_\nu \Lambda_\mu$, the field equation for the vector sector of the linearized gravity comes only from the gauge-fixing term~\eqref{eq:new-gf-term}. This equation reads
\beq
\nabla_\mu \nabla^\rho h^{(V)}_{\nu\rho} + \nabla_\nu \nabla^\rho h^{(V)}_{\mu\rho} = 0\eqend{,}
\eeq
because $h^{(V)\alpha}_\alpha = 2 \nabla_\alpha V^\alpha = 0$.  Notice that this equation is independent of the gauge parameter $\alpha$. It can be written as
\beq
\nabla^\rho h^{(V)}_{\nu\rho} \propto \xi_\nu\eqend{,}  \label{eq:Killing-vector-mode}
\eeq
where $\xi_\nu$ is a Killing vector.

The part of the field $h^{(V)}_{\mu\nu}$ relevant to the charge $Q^{(H)}_{(\sigma,\rR)}$ corresponding to the rotation Killing vector $\xi_{(\sigma,\rR)}^\mu$~(\ref{eq:rotation-Killing}), which will be given as
\beq
h^{(\sigma,\rR)}_{\mu\nu} = \nabla_\mu W^{(\sigma,\rR)}_{\nu} + \nabla_\nu W^{(\sigma,\rR)}_{\mu}\eqend{,}
\label{eq:h-sigma-rR}
\eeq
can be obtained by postulating 
\beq
W^{(\sigma,\rR)\mu} = F^{(\sigma,\rR)}(t)\xi_{(\sigma,\rR)}^\mu\eqend{,} \label{eq:V-sigma-rR}
\eeq
and solving 
Eq.~\eqref{eq:Killing-vector-mode}.  Here, $F^{(\sigma,\rR)}(t)$ is a time-dependent operator.  We find
\beq
F^{(\sigma,\rR)}(t) = \alpha a_{(\sigma,\rR)}f_1(t) + b_{(\sigma,\rR)}f_2(t)\eqend{,} \label{eq:ambiguous}
\eeq
where $a_{(\sigma,\rR)}$ and $b_{(\sigma,\rR)}$ are constant Hermitian operators, 
\begin{subequations}
\beqa
\dot{f}_1(t)  & = & \frac{1}{\cosh^{n+1}t}\int_0^t \cosh^{n+1}\tau\,d\tau\eqend{,} \label{eq:f1-definition}\\
\dot{f}_2(t) & = & \frac{1}{\cosh^{n+1}t}\eqend{,} \label{eq:f2-definition}
\eeqa
\end{subequations}
with $\dot{f}_i(t) = df_i(t)/dt$, $i=1,2$.\footnote{The $\alpha$-dependence of Eq.~(\ref{eq:ambiguous}) was chosen so that the commutator~(\ref{eq:a_b_commutator}) found from the symplectic product between the modes $f_1(t)\xi^\mu_{(\sigma,\mathrm{R})}$ and $f_2(t)\xi^\mu_{(\sigma,\mathrm{R})}$ is $\alpha$-independent. The operator $F^{(\sigma,\rR)}(t)$ is defined only up to addition of a constant operator. Note that addition of a constant to $F^{(\sigma,\rR)}(t)$ does not alter the field $h^{(\sigma,\rR)}_{\mu\nu}$.}

Recalling that $B^\mu = - \alpha^{-1}\nabla_\nu H^{\mu\nu}$, the $\alpha\to 0$ limit yields
\begin{subequations}
\beqa
B^{(\sigma,\rR)\mu} & = & - \lim_{\alpha\to 0}\frac{1}{\alpha}\nabla_\nu h^{(\sigma,\rR)\mu\nu}
= a_{(\sigma,\rR)}\xi^\mu_{(\sigma,\rR)}\eqend{,}\nonumber \\ \\
Q_{(\sigma,\rR)}^{(H)} & = & \lim_{\alpha\to 0}\int_{\Sigma}d\Sigma\,
n^\mu \xi^\nu h_{\mu\nu}^{(\sigma,\rR)} = b_{(\sigma,\rR)}\eqend{.}
\eeqa
\end{subequations}
Thus, from Eq.~\eqref{eq:V-decomposition} we find $a_{(\sigma,\rR)} = B_{(0)}^{(\sigma,\rR)}$, while Eq.~\eqref{pAandQA} leads to $b_{(\sigma,\rR)} = p_{(\sigma,\rR)}$, the canonical momentum conjugate to $B_{(0)}^{(\sigma,\rR)}$.  That is,
\beq
\label{eq:a_b_commutator}
[a_{(\sigma,\rR)},b_{(\sigma',\rR)}] = i\delta_{\sigma\sigma'}\eqend{.}
\eeq

Now, the de~Sitter-invariant Bunch-Davies vacuum state $|0\rangle$ is annihilated by the operator $A_{(\sigma,\rR)}$, i.e.\ we have $A_{(\sigma,\rR)}|0\rangle=0$, where $A_{(\sigma,\rR)}$ is a linear combination of $a_{(\sigma,\rR)}$ and $b_{(\sigma,\rR)}$. The operator $F^{(\sigma,\rR)}(t)$ in Eq.~\eqref{eq:ambiguous} is then expressed as
\beqa
F^{(\sigma,\rR)}(t)  & = &
\left[f_1(t) + ic_0 f_2(t)\right]A_{(\sigma,\rR)}\nonumber \\
&&  + \left[ f_1(t) - i c_0 f_2(t)\right]A^\dagger_{(\sigma,\rR)}\eqend{.}
\label{eq:F-sigma-rR-alternative}
\eeqa
That is, the function $f_1(t) + ic_0 f_2(t)$ corresponds to the positive-frequency mode for the Bunch-Davies vacuum
state $|0\rangle$.  The constant $c_0$ can be found as in Ref.~\cite{Faizal:2011iv}, and it reads  
\beq
c_0 = \frac{\sqrt{\pi}\Gamma(\frac{n+2}{2})}{2\Gamma(\frac{n+3}{2})}\eqend{.}
\label{eq:c0-def}
\eeq
The exact value of this constant is not important; what matters is that Eq.~\eqref{eq:F-sigma-rR-alternative} is independent of $\alpha$.  

Finally, the comparison between Eqs.~\eqref{eq:ambiguous} and~\eqref{eq:F-sigma-rR-alternative} shows that
\begin{subequations}
\beqa
B_{(0)}^{(\sigma,\rR)} & = & \frac{1}{\alpha}\left(A_{(\sigma,\rR)} + A^\dagger_{(\sigma,\rR)}\right)\eqend{,}\\
Q^{(H)}_{(\sigma,\rR)} & = & ic_0 \left(A_{(\sigma,\rR)}- A_{(\sigma,\rR)}^\dagger\right)\eqend{,}
\eeqa
\end{subequations}
and
\beq
\left[ A_{(\sigma,\rR)},A^\dagger_{(\sigma,\rR)}\right] = - \frac{\alpha}{2c_0}\eqend{.}
\eeq
Thus, we find
\beq
\langle 0|Q^{(H)}_{(\sigma,\rR)}Q^{(H)}_{(\sigma',\rR)}|0\rangle
= - \frac{\alpha c_0}{2} \to 0\ \ \textrm{as}\ \ \alpha \to 0\eqend{.}
\eeq
That is, $\langle 0| \omega Q^{(H)}_{(\sigma,\rR)}|0\rangle = 0$ for all canonical variables $\omega$ except
for $\omega = B_{(0)}^{(\sigma,\rR)}$.

We now turn to the conserved charge $Q^{(H)}_{(\sigma,\rB)}$ associated with the boost Killing vectors~(\ref{eq:boost-Killing}). The relevant part of $h_{\mu\nu}^{(V)}$ to this charge is denoted by $h_{\mu\nu}^{(\sigma,\rB)}$ and reads
\beq
h^{(\sigma,\rB)}_{\mu\nu} = \nabla_\mu W^{(\sigma,\rB)}_{\nu} + \nabla_\nu W^{(\sigma,\rB)}_{\mu}\eqend{,}
\label{eq:h-sigma-rB}
\eeq
where
\begin{subequations}
\beqa
W^{(\sigma,\rB)}_0 & = & F^{(\sigma,\rB)}(t) Y_{(1\sigma)}\eqend{,}\\
W^{(\sigma,\rB)}_i & = & - \frac{\cosh^2 t}{n-1}\left[ \dot{F}^{(\sigma,\rB)}(t) \right. \nonumber \\
&& \left. + (n-1)\tanh t 
F^{(\sigma,\rB)}(t)\right] \mathcal{D}_i Y_{(1\sigma)}\eqend{.}
\eeqa
\end{subequations}
The operator $F^{(\sigma,\rB)}(t)$ is again given as
\beq
F^{(\sigma,\rB)}(t) = \alpha a_{(\sigma,\rB)}f_1(t) + b_{(\sigma,\rB)}f_2(t)\eqend{,}
\eeq
where $f_1(t)$ and $f_2(t)$ are given by Eqs.~\eqref{eq:f1-definition} and \eqref{eq:f2-definition}, respectively, and $a_{(\sigma,\rB)}$ and $b_{(\sigma,\rB)}$ are constant operators.  

Following the same procedure as in the case for the rotation Killing vector, we find $a_{(\sigma,\rB)} = B_{(0)}^{(\sigma,\rB)}$ and $b_{(\sigma,\rB)} = - Q^{(H)}_{(\sigma,\rB)}/2$ in the Landau-gauge limit.  The operator $F^{(\sigma,\rB)}(t)$ is expressed in terms of the annihilation and creation operators, $A_{(\sigma,\rB)}$ and $A_{(\sigma,\rB)}^\dagger$, with $A_{(\sigma,\rB)}|0\rangle =0$, in exactly the same way as in Eq.~\eqref{eq:F-sigma-rR-alternative}. Hence, we find
\begin{subequations}
\beqa
B_{(0)}^{(\sigma,\rB)} & = & \frac{1}{\alpha}\left(A_{(\sigma,\rB)} + A^\dagger_{(\sigma,\rB)}\right)\eqend{,}\\
Q^{(H)}_{(\sigma,\rB)} & = & -2ic_0 \left(A_{(\sigma,\rB)}- A_{(\sigma,\rB)}^\dagger\right)\eqend{,}
\eeqa
\end{subequations}
and
\beq
\left[ A_{(\sigma,\rB)},A^\dagger_{(\sigma,\rB)}\right] = \frac{\alpha}{c_0}\eqend{.}
\eeq
Then, we again have $\langle 0| Q^{(H)}_{(\sigma,\rB)}Q^{(H)}_{(\sigma',\rB)}|0\rangle = 0$ in the Landau-gauge limit.  Thus, $\langle 0|\omega Q^{(H)}_{(\sigma,\rB)}|0\rangle = 0$ for all canonical variable $\omega$ except for $\omega = B_{(0)}^{(\sigma,\rB)}$.

\section{Some Details of the Calculations for the Ghost Fields in Sec.~\ref{sec:Killing-vector-modes}}                                              %
\label{app-FPscalar-no-contrib}                                                                                                                     %

In this section we provide some details of the calculations  in Sec.~\ref{sec:Killing-vector-modes}, where it is shown that the linearized ghost charges $Q^{(c)}_A$ and $Q^{(\cbar)}_A$ annihilate the tree-level vacuum state. We show first that the scalar sectors of the FP ghosts do not contribute to the tree-level charges.

The scalar sector of the field $c^\mu$ is given, at tree level, by $\nabla^\mu \Phi$, where the field $\Phi$ satisfies Eq.~\eqref{eq:scalar-sector-ghost}. Then the contribution of this sector to the charge is
\beqa
Q_A^{(c,S)} & = & 2\int_\Sigma d\Sigma\, n^\mu \xi_A^\nu \left[\nabla_\mu \nabla_\nu \Phi -
\left( 1 + \frac{1}{\beta}\right) g_{\mu\nu}\Box\Phi\right] \nonumber \\
& = & 2\int_\Sigma d\Sigma\,
n_\mu 
\nabla_\nu\left[ \xi_A^\nu \nabla^\mu\Phi - \xi_A^\mu \nabla^\nu\Phi
+ (\nabla^\nu\xi_A^\mu)\Phi\right] \nonumber \\
&& + m^2 \int_\Sigma d\Sigma\, n_\mu \xi_A^\mu \Phi \nonumber \\
& = & m^2 \int_\Sigma d\Sigma\, n_\mu \xi_A^\mu\Phi\eqend{,}
\eeqa
where we have used Eq.~\eqref{eq:scalar-sector-ghost}, the equation $\Box\xi_A^\mu = - (n-1)\xi_A^\mu$ and the generalized Stokes theorem.
Hence the contribution of the scalar sector to the conserved FP-ghost charge at tree level vanishes in the limit $m\to 0$.

Next we derive Eq.~\eqref{eq:interesting}. First we note, using $(d/dz)F(a,b;c;z) = (ab/c)F(a+1,b+1;c+1;z)$, that
\begin{splitequation}
&\frac{d\ }{dt}\phantom{}_2F_1\left(b_1^+,b_1^-; \frac{n+2}{2};\frac{1-i\sinh t}{2}\right) \\
&= - i\frac{b_1^+b_1^-}{n+2}\cosh t\,
\phantom{}_2F_1\left(b_1^++1,b_1^-+1; \frac{n+4}{2};\frac{1-i\sinh t}{2}\right) \\
&\approx - m^2\frac{i\cosh t}{n+2} \phantom{}_2F_1\left(n+2,1; \frac{n+4}{2}; \frac{1-i\sinh t}{2}\right)\eqend{,}
\end{splitequation}
for small $m$. Then the formula~\cite{gradshteyn2007}
\beqa
&& \phantom{}_2F_1\left(2a, 2b; a + b + \frac{1}{2}; \frac{1 - \sqrt{z}}{2}\right) \nonumber \\
&& = A\phantom{}_2F_1\left(a, b; \frac{1}{2}; z\right) + B\sqrt{z}\, \phantom{}_2F_1\left(a + \frac{1}{2}, b 
+ \frac{1}{2}; \frac{3}{2}; z\right)\eqend{,} \nonumber \\
\eeqa
with the constants
\begin{subequations}
\beqa
 A  & = & \frac{\sqrt{\pi}\Gamma\left(a + b + \frac{1}{2}\right)}{\Gamma\left(a + \frac{1}{2}\right)\Gamma\left(b + \frac{1}{2}\right)}\eqend{,}\\
B &  = & -\frac{2\sqrt{\pi}\Gamma\left(a + b + \frac{1}{2}\right)}{\Gamma(a) \Gamma(b)}\eqend{,}
\eeqa
\end{subequations}
allows us to write
\beqa
&& \frac{d\ }{dt}\phantom{}_2F_1\left(b_1^+,b_1^-; \frac{n+2}{2};\frac{1-i\sinh t}{2}\right) \nonumber \\
&& \approx -i c_0 m^2 \cosh t\, \phantom{}_2F_1\left( \frac{n+2}{2},\frac{1}{2}; \frac{1}{2}; - \sinh^2 t\right)\nonumber \\
&& - m^2 \cosh t\sinh t\, \phantom{}_2F_1\left( \frac{n+3}{2},1; \frac{3}{2}; - \sinh^2 t\right)\nonumber\\
&& = - \frac{m^2}{\cosh^{n+1}t} \left( ic_0 + \int_0^t \cosh^{n+1}\tau\,d\tau\right)\eqend{,}
\eeqa
where the constant $c_0$ is given by Eq.~(\ref{eq:c0-def}).  We have used
\beq
\phantom{}_2F_1(a,b;c;z) = (1-z)^{c-a-b}\phantom{}_2F_1(c-a,c-b;c;z)\eqend{,}
\eeq
to find
\beqa
&& \phantom{}_2F_1\left( \frac{n+3}{2},1;\frac{3}{2};-\sinh^2 t\right)\nonumber \\
&&  = \frac{1}{\sinh t (\cosh t)^{n+2}}\int_0^t (\cosh \tau)^{n+1}\,d\tau\eqend{.}
\eeqa

\section{Equivalence of Hamiltonian and Lagrangian Perturbation Theories for Scalar QED}                                                            %
\label{app-scalarQED}                                                                                                                               %

The Lagrangian density for QED with charged scalar field $\phi$ in Minkowski spacetime is
\beqa
\mathcal{L} & = &
- (\partial^\mu \phi^\dagger + i e A^\mu \phi^\dagger)(\partial_\mu \phi - i e A_\mu\phi) - m^2 
\phi^\dagger\phi\nonumber \\
&&  - \frac{1}{4}F_{\mu\nu}F^{\mu\nu} - \frac{1}{2\alpha}(\partial_\mu A^\mu)^2 \eqend{,}
\eeqa
where $A_\mu$ is the gauge potential and $F_{\mu\nu} = \partial_\mu A_\nu - \partial_\nu A_\mu$.  The
interaction Lagrangian density consists of the nonquadratic terms in the Lagrangian density:
\beq
\mathcal{L}_I = ie A^\mu (\phi\partial_\mu\phi^\dagger - \phi^\dagger\partial_\mu\phi) - 
e^2A^\mu A_\mu\phi^\dagger\phi
\eqend{.}
\eeq 

The canonical momentum density conjugate to $\phi^\dagger$ is
\beq
\pi  = \dot{\phi} - ieA^0\phi\eqend{,}
\eeq
and the canonical momentum density conjugate to $\phi$ is $\pi^\dagger$.  
The interaction Hamiltonian density, i.e.\ the
nonquadratic part of the Hamiltonian density is,
\beqa
\mathcal{H}_I & = & - ie A^0 (\phi\pi^\dagger - \phi^\dagger\pi) 
- ie A^i (\phi\partial_i\phi^\dagger - \phi^\dagger\partial_i \phi) \nonumber \\
&& + e^2 A^i A_i \phi^\dagger\phi\eqend{.}
\eeqa
In both Lagrangian and Hamiltonian perturbation theories in the interaction picture, the field operators satisfy the free-field equations.  Thus, we have $\pi = \dot{\phi}$.  This allows a direct comparison between the interaction Lagrangian and Hamiltonian densities as
\beq
\mathcal{L}_I = - \mathcal{H}_I  - e^2 A^0 A_0 \phi^\dagger\phi\eqend{.}
\eeq
Thus, $\mathcal{L}_I \neq - \mathcal{H}_I$.

The difference between $\mathcal{L}_I$ and $-\mathcal{H}_I$ is accounted for by the fact that in Lagrangian perturbation theory the time derivatives are applied to the propagator as follows [with $x=(t,\mathbf{x})$ and $x'=(t',\mathbf{x}')$]:
\beqa
&& \partial_t \partial_{t'} T \langle 0|\phi(x)\phi^\dagger(x')|0\rangle \nonumber \\
&& =  T\langle 0|\dot{\phi}(x)\dot{\phi}^\dagger(x')|0\rangle 
+ \delta(t-t') \langle 0| \left[ \phi(x),\phi^\dagger(x')\right]|0\rangle\nonumber \\
& & =  \langle 0|\pi(x)\pi^\dagger(x')|0\rangle + i \delta^{(4)}(x-x')\eqend{.}
\eeqa
Thus, in Lagrangian perturbation theory there is an extra interaction term $-e^2 A^0 A_0 \phi^\dagger\phi$ and the field $\pi = \dot{\phi}$ is replaced by $\pi + \gamma$, where $\gamma$ has the propagator
\beq
T\langle 0|\gamma(x)\gamma^\dagger(x')|0\rangle = i \delta^{(4)}(x-x')\eqend{.}
\eeq
Integrating out the fictitious field $\gamma(x)$ generates the following effective interaction term:
\beqa
\Delta \mathcal{L}_I & = & - i\int d^4x'\, e A^0(x)\phi^\dagger(x)
T\langle 0| \gamma(x)\gamma^\dagger(x')|0\rangle\nonumber \\
&& \times  \left[ - e\phi(x')A^0(x')\right] \nonumber \\
& = & e^2 A^0(x)A_0(x) \phi^\dagger(x)\phi(x)\eqend{.}
\eeqa
Thus, we have
\beq
\mathcal{L} + \Delta \mathcal{L}_I = - \mathcal{H}_I\eqend{,}
\eeq
which shows that Lagrangian and Hamiltonian perturbation theories are equivalent in QED with charged scalar field.

\bibliography{refs_ghl_v6}

\end{document}